\documentclass[preprint2]{aastex}
\usepackage{amssymb,latexsym,graphics,eufrak}
\begin{document}
\title{Structure of pair winds from compact objects
with application
to emission from hot bare strange stars}
\author{A.G. Aksenov~\altaffilmark{1}, M.~Milgrom, V.V.~Usov}
\affil{Department of Condensed Matter Physics, Weizmann Institute,
Rehovot 76100, Israel} \altaffiltext{1}{Institute of
Theoretical and Experimental Physics, B.~Cheremushkinskaya, 25,
Moscow 117259, Russia; Alexei.Aksenov@itep.ru}
\begin{abstract}
We consider a stationary, spherically outflowing wind consisting
of electron-positron pairs and photons. We do not assume thermal
equilibrium, and include the two-body processes that occur in such
a wind: M{\o}ller and Bhaba scattering of pairs, Compton
scattering, two-photon pair annihilation, and two-photon pair
production, together with their radiative three-body variants:
bremsstrahlung, double Compton scattering, and three-photon pair
annihilation, with their inverse processes. In the concrete
example described here, the wind injection source is a hot, bare,
strange star. Such stars are thought to be powerful sources of
pairs created by the Coulomb barrier at the quark surface. We
present a new, finite-difference scheme for solving the
relativistic kinetic Boltzmann equations for pairs and photons.
Using this method we study the kinetics of the wind particles and
the emerging emission for total luminosities of $L =
10^{34}-10^{42}$ ergs~s$^{-1}$ (the upper limit being set, at the
moment, by computational limitations). We find the rates of
particle number and energy outflows, outflow velocities, number
densities, energy spectra, and other parameters for both photons
and pairs as functions of the distance. We find that for
$L>2\times 10^{35}$ ergs~s$^{-1}$, photons dominate the emerging
emission. For all values of $L$ the number  rate of emerging pairs
is bounded: $\dot N_e\lesssim \dot N_e^{\rm max}\simeq 10^{43}$
s$^{-1}$. As $L$ increases from $\sim 10^{34}$ to $10^{42}$
ergs~s$^{-1}$, the mean energy of emergent photons decreases from $\sim
400-500$ keV to 40 keV, as the spectrum changes in shape from that
of a wide annihilation line to nearly a blackbody spectrum with a
high energy ($> 100$ keV) tail. These results are pertinent to the
deduction of the outside appearance of hot bare strange stars,
which might help discern them from neutron stars.

\end{abstract}
\keywords{radiation mechanisms: thermal --- plasmas ---
X-rays: stars --- radiative transfer ---
stars: neutron}
\section{Introduction}
There is now compelling evidence that electron-positron ($e^\pm$)
pairs form and flow away from many compact astronomical objects
identified as neutron stars and black holes. Among these objects
are pulsars (Sturrock 1971; Ruderman \& Sutherland 1975, Arons 1981;
Usov \& Melrose 1996; Baring \& Harding 2001), accretion-disc coronas
of the Galactic X-ray binaries (White \& Lightman 1989;
Sunyaev
et al. 1992), active galactic nuclei (Jelley 1966; Herterich 1974;
Begelman, Blandford, \& Rees 1984; Lightman \& Zdziarski 1987;
Sikora 1994; Blandford \& Levinson 1995; Reynolds et al. 1996;
Yamasaki, Takahara, \& Kusunose 1999), cosmological $\gamma$-ray
bursters (Eichler et al. 1989;
Paczy\'nski 1990;
Usov 1992), etc. Several pair-production mechanisms are
thought to operate in these objects. In pulsar magnetospheres, it is
supposedly effected via conversion of single $\gamma$-photons into
$e^\pm$ pairs in a strong magnetic field ($\gamma B\rightarrow
e^+ e^-  B$), while in galactic nuclei, the
pairs are created mostly in photon-photon collisions
($\gamma \gamma\rightarrow e^+ e^-$).
Thermal neutrinos and antineutrinos radiated by a
very hot compact object may be absorbed near its surface
and produce $e^\pm$ pairs ($\nu \tilde\nu\rightarrow e^+ e^-$).
For example in very young (age $< 10$~s), compact objects
with mean internal temperatures of $\sim 10^{11}$
K, the pair
luminosity  produced via this mechanism
may be as high as
$\sim 10^{51}$
ergs~s$^{-1}$ (e.g., Eichler et al. 1989; Haensel,
Paczy\'nski, \& Amsterdamski 1991), which is a typical
luminosity of cosmological $\gamma$-ray bursters. Such pair winds are,
therefore, discussed as possible sources of
cosmological $\gamma$-ray
bursts
(for a review, see Piran 2000). Recently it was shown that
strange
stars, which are made entirely of deconfined quarks (e.g.,
Witten 1984; Alcock, Farhi, \& Olinto 1986a; Haensel, Zdunik, \&
Schaeffer 1986), may also be powerful sources of $e^\pm$ pairs
(Usov 1998, 2001a). In this case, the pairs are created in an extremely
strong electric field at the quark surface. The thermal luminosity
of a bare strange star in $e^\pm$ pairs depends on the surface
temperature $T_{\rm S}$ and may be as high as $\sim 10^{50}-10^{51}$
ergs~s$^{-1}$ at the moment of formation when
$T_{\rm S}$ may be up to $\sim 10^{11}$~K (see Fig.~1).
The
luminosity of a young bare strange star in pairs may remain high enough
($\gtrsim 10^{36}$ ergs~s$^{-1}$) for $\sim10^3$ yr (Page \& Usov 2002).

The properties of pair plasmas have been studied extensively, but
most of the initial attention focused on plasmas in thermal
equilibrium (e.g., Bisnovatyi-Kogan, Zeldovich, \& Sunyaev 1971;
Lightman 1981; Svensson 1982; Guilbert \& Stepney 1985; Zdziarski
1985 and references therein). It is conceivable that such equilibrated
plasmas exist in some objects. Very powerful pair winds are among
these
objects. For an energy injection rate of
$\dot {\rm E}\gtrsim 10^{50}$
ergs~s$^{-1}$, the pair density near a
compact object of radius
$\sim 10^6$ cm is very high, and the
outflowing pairs and photons are nearly in thermal
equilibrium almost up to the wind photosphere (e.g., Paczy\'nski
1990). The outflow process of such a wind may be described fairly
well by relativistic hydrodynamics (Paczy\'nski 1986, 1990; Goodman
1986; Grimsrud \& Wasserman 1998; Iwamoto \& Takahara 2002). The
emerging emission consists mostly of photons, so $L_\gamma\simeq
\dot {\rm E}$. The photon spectrum is roughly a blackbody with a
temperature of $\sim 10^{10}(\dot {\rm E}/10^{50}~{\rm
ergs~s}^{-1})^{1/4}$~K. The emerging luminosity in $e^\pm$ pairs
is very small, $L_e=\dot {\rm E}-L_\gamma \sim 10^{-6}
L_\gamma\ll L_\gamma$. All this applies
roughly down to $\dot {\rm E} \sim 10^{42}-10^{43}$ ergs~s$^{-1}$.
For $\dot {\rm E} < 10^{42}$ ergs~s$^{-1}$,
which is the region we explore, the thermalization time for the
$e^\pm$ pairs and photons is longer than the escape time, and pairs
and photons are not in thermal equilibrium (see below).
Recently, a brief account of the emerging emission from such
a pair wind
has been given by Aksenov, Milgrom, \& Usov (2003).
In the present paper, we study numerically the kinetics of $e^\pm$
pairs and photons in a stationary pair wind with
energy injection rates $\dot {\rm E}\leq 10^{42}$ ergs~s$^{-1}$, and
find the structure of the wind and
its emerging emission in $e^\pm$ pairs and photons. We
assume that the outflowing wind consist of $e^\pm$ pairs and
photons only. For the sake of concreteness, we
take, in the present study, the input pair injection at the inner
surface
to correspond to a hot bare strange star. We use the
Boltzmann equations
for both $e^\pm$
pairs and photons (kinetic theory approach). The principal
advantage of this approach compared to the Monte Carlo method
is that it gives better photon statistics. Besides, the
kinetic theory approach may still be used when the optical thickness
is $\gg 1$, where the Monte Carlo method is ineffective.
Some examples of the kinetic theory approach for investigations
of non-equilibrium pair plasmas can be found in Fabian et al.
(1986), Ghisellini (1987), Svensson (1987), and Coppi (1992).
The remainder of the paper is organized as follows.
In \S~2 we formulate the equations that describe the pair wind
from a hot bare strange star, and the boundary conditions. In \S~3
we describe the computational method used to solve these equations.
In \S~4 we present the results of our numerical simulations. Finally,
in \S~5, we discuss some astrophysical applications.

\section{Formulation of the problem}
We consider, then, an $e^\pm$ pair wind that flows away from a hot,
bare,
unmagnetized, strange star with
a radius of $R=10^6$ cm.
We assume that the temperature is the same
over the surface of
the star. The pair flux $F_e$
from the unit surface
of strange quark matter (SQM) depends
on the surface temperature
$T_{\rm S}$ only (Usov 1998, 2001a); and the pair wind is spherical.
At a distance $r$ from the stellar center and at time
$t$, the state of the plasma may be described by the distribution
functions $f_{e^\pm} ({\bf p}, r,t)$ and $f_\gamma
({\bf p}, r,t)$
for positrons $(+)$, electrons $(-)$, and
photons, respectively,
where ${\bf p}$ is the momentum of particles.
There is no emission of
nuclei
from the stellar surface, and therefore, the distribution
functions of positrons and electrons are identical,
$f_{e^+}=f_{e^-}=f_e/2$.

In the first run of our simulations the outflow of $e^\pm$ pairs from
the stellar surface starts at
the moment $t=0$, and its rate is
constant with time ($\dot {\rm E}=10^{34}$ ergs~s$^{-1}$) till
the moment when a steady state is established in the examined space
domain (see below). Then, the rate of the pair outflow increases
instantly by a factor of 10, and the second run
starts and continues till a new steady state
is established, and so on.
So, we find the structure of stationary $e^\pm$ pair winds and their
emerging radiation for $\dot {\rm E}=10^{34},\,10^{35},\, ...\,,\,
10^{42}$ ergs~s$^{-1}$.

\subsection{Equations}

We use the relativistic Boltzmann
equations for
the $e^\pm$ pairs and photons, whereby the distribution function
for the particles of type $i$, $f_i
(|{\bf p}|, \mu, r, t)$, ($i=e$ for
$e^\pm$ pairs and $i=\gamma$ for photons), satisfies (e.g.,
de Groot,
van Leeuwen, \& van Weert 1980; Mihalas 1984;
Mezzacappa \& Bruenn 1993)
\begin{equation}
  \frac{1}{c}\frac{\partial f_i}{\partial t}
 +\beta_i\left( \mu\frac{\partial f_i}{\partial r}
         +\frac{1-\mu^2}{r}\frac{\partial f_i}{\partial\mu}
   \right)
 =\sum_q\left[\tilde\eta_i^q-\chi_i^q f_i\right], \label{dfi}
\end{equation}
where $\mu =\cos \theta$, $\theta$ is the
angle between the radius-vector ${\bf r}$ from the stellar center
and the particle
momentum ${\bf p}$, $p=|{\bf p}|$,
and
\begin{equation}
\beta_e=v_e/c=[1-(m_e c^2/\epsilon_e)^2]^{1/2},\,\,\,\,\,\,\,
\beta_\gamma =1\,, \label{betae}
\end{equation}
while
\begin{equation}
\epsilon_e=c[p^2+(m_ec)^2]^{1/2}\,\,\,\,\,\,\,{\rm and}\,\,\,\,\,\,\,
\epsilon_\gamma=pc
\label{epsilone}
\end{equation}
are the energy of electrons and photons, respectively. Also
$\tilde\eta_i^q$ is the emission coefficient for the production of
a particle of type $i$ via the physical process labeled by $q$,
and $\chi_i^q$ is the corresponding absorption coefficient. The
summation runs over all considered physical processes that involve
a particle of type $i$. Gravity is neglected (but its effects are briefly
discussed in \S 5). This is a good approximation for $\dot{\rm E}>
\tilde L_{\rm Edd}$, while for $\dot{\rm E}\lesssim\tilde L_{\rm Edd}$ the
gravity effects may be important, where
\begin{equation}
\tilde L_{\rm Edd}\simeq 7\times 10^{34} (M/M_\odot)~{\rm ergs~s}^{-1}
\,,
\end{equation}
is the Eddington luminosity for the pair
plasma,
above which radiation-pressure forces dominate over
gravity.

The particle densities are given by
\begin{equation}
n_i(r,t)=\int f_i ({\bf p}, r,t)d{\bf p}\,. \label{ni}
\end{equation}

For convenience of numerical simulations we use, instead of $f_i$,
the quantities
\begin{equation}
E_i(\epsilon_i,\mu,r,t)
   ={2\pi  \epsilon_i^3\beta_i f_i\over c^3},
\label{Ei}
\end{equation}
as are used in the ``conservative'' numerical method, which can
provide exact conservation of energy on a finite
computational grid (see below).
Since
\begin{equation}
    \epsilon_i
    f_i d\mathbf{r}d\mathbf{p}
   ={2\pi
\epsilon_i^3 \beta_i f_i\over c^3}
    d\mathbf{r}d\epsilon_i d\mu= E_i d\mathbf{r}d\epsilon_i d\mu\,,
    \label{rp}
\end{equation}
we see that $E_i$
is the energy density in the
$\{{\bf r},\mu, \epsilon_i, \}$ phase space,
in which the volume element
is $d\mathbf{r}d\epsilon_i d\mu=4\pi r^2dr d\epsilon_i d\mu$.

From equations (\ref{dfi}) and (\ref{Ei}) the Boltzmann equations can be
written in terms of $E_i$
as
\begin{eqnarray}
  \frac{1}{c}\frac{\partial E_i}{\partial t}
 +{\mu\over r^2}\frac{\partial}{\partial r}
 (r^2 \beta_i E_i)
 +\frac{1}{r}\frac{\partial
 }{\partial\mu}
\left[(1-\mu^2)\beta_i E_i\right]
 \nonumber \\
 =\sum_q\left[\eta_i^q-\chi_i^q E_i\right], \label{Boltzmann}
\end{eqnarray}
where
\begin{equation}
\eta_i^q={2\pi \epsilon_i^3\beta_i\tilde\eta_i^q\over c^3} \label{etai}
\end{equation}
This form of the Boltzmann equations is said to be
``conservative'', in the case without gravity, because the
``transport'' terms can be written as a derivative of a flux
divided by a volume (Harleston \& Holcomb 1991). This allows one
to develop a conservative finite difference code for numerical
simulations and to carry out calculations with large time steps
even for very large optical depths. In our simulations
we use the
Boltzmann equations (\ref{Boltzmann}) for both
pairs and
photons.

\subsection{Boundary conditions}
The computational boundaries are
\begin{equation}
r_{\rm int} < r < r_{\rm ext}, \,\,\,\,\,\,\,0 < t < t_{\rm st}\,,
\label{rint}
\end{equation}
where $r_{\rm ext}=1.65\times 10^8$ cm is chosen as the radius of the
external boundary, and $t_{\rm st}$ is the time when the wind
approaches stationary closely enough. The
internal radius is that typical of a strange star ($r_{\rm
int}=R$). At this radius, the injection rate of pairs
is taken as (Usov 1998, 2001a)
\begin{equation}
\dot N_e^{\rm in}=4\pi R^2F_e\,,
\end{equation}
where
\begin{eqnarray}
F_e=3\times 10^{42}
             \exp\,(- 0.593\zeta)
             \nonumber \\
             \times
             \left[\frac{\displaystyle\ln(1+2\zeta^{-1})}
                    {\displaystyle(1+0.074\zeta)^3}
+
                    \frac{\displaystyle\pi^5\zeta}
                    {\displaystyle2(13.9+\zeta)^4}
             \right]\,{\rm cm}^{-2}{\rm s}^{-1},
\label{Fpm}
\end{eqnarray}
and $\zeta=20(T_{\rm S}/10^9\,{\rm K})^{-1}$.
The energy
spectrum of injected pairs
is thermal with temperature $T_{\rm S}$, and their
angular distribution is isotropic for $0\leq \mu \leq 1$.

The
energy injection rate is equal to the thermal luminosity of the
strange star in $e^\pm$ pairs,
\begin{equation}
\dot {\rm E}=
L_{\rm SS}=
\dot N_e^{\rm in}[m_ec^2 +(3/2)k_{\rm B}T_{\rm S}] \,,
\label{dotE}
\end{equation}
where $k_{\rm B}$ is the Boltzmann constant (Usov 2001a). For the
range of the energy injection rates we consider, $\dot {\rm E}=
10^{34} - 10^{42}$ ergs~s$^{-1}$, the surface temperature is in a
rather narrow range, $T_{\rm S}\simeq 4\times 10^8- 10^9$~K (see
Fig.~1).

The thermal emission of photons from the surface of a bare strange
star is strongly suppressed if the surface temperature is not very
high, $T_{\rm S}\ll 10^{11}$ K (Alcock et al. 1986a). The reason
is that the plasma frequency of quarks in SQM is very large,
$\hbar \omega_{\rm p}\simeq 20-25$ MeV, and only hard photons with
energies $\epsilon_\gamma>\hbar \omega_{\rm p}$ can propagate in
SQM. The luminosity in such hard photons, which are in
thermodynamic equilibrium with quarks, decreases very fast for
$T_{\rm S} \ll \hbar \omega_{\rm p}/k_{\rm B}$ (Chmaj, Haensel, \&
Slomi\'nski 1991; Usov 2001a), and in our case when $T_{\rm S}$ is
$\lesssim 10^9$ K, it is negligible (see Fig. 1). However,
low-energy ($\epsilon_\gamma <\hbar \omega_{\rm p}$) photons may
leave SQM if they are produced by a nonequilibrium process in the
surface layer of thickness of $\sim c/\omega_{\rm p}\sim 10^{-12}$
cm (Chmaj et al. 1991). Recently, the emissivity of SQM in
nonequilibrium quark-quark bremsstrahlung radiation has been
estimated by Cheng \& Harko (2003), taking into account the
Landau-Pomeranchuk-Migdal effect, and it is shown that the SQM
emission in nonequilibrium photons is suppressed at least by a
factor of $10^6$ in comparison with blackbody emission (see Fig.
1). It was argued that SQM is a color superconductor with a very
high critical temperature $T_c\sim 10^{11}-10^{12}$ K (Alford,
Rajagopal, \& Wilczek 1999; Alford, Bowers, \& Rajagopal 2001). In
this case, the nonequilibrium emission is suppressed further. In
our simulations, we neglect the surface emission of photons
altogether.

The stellar surface is assumed to be a perfect mirror
for both $e^\pm$ pairs and photons.
At the external boundary ($r=r_{\rm ext}$), the pairs and photons
escape freely from the studied region, i.e., the inward ($\mu <
0$) flux of both $e^\pm$ pairs and photons vanishes.

\subsection{Physical processes in the pair plasma}
Although the pair plasma ejected from the strange-star surface
contains no radiation, as the plasma moves outwards photons are
produced by pair annihilation ($e^+e^-\rightarrow \gamma\gamma$).
Other two-body processes that occur in the outflowing plasma are
M{\o}ller ($e^+e^+ \rightarrow e^+e^+ $ and $e^-e^-\rightarrow
e^-e^-$) and Bhaba ($e^+e^-\rightarrow e^+e^-$) scattering,
Compton scattering ($\gamma e\rightarrow \gamma e$), and
photon-photon pair production ($\gamma\gamma\rightarrow e^+e^-$).
Two-body processes do not change the total number of particles in
a
system, and thus cannot, in themselves, lead to thermal
equilibrium.
For this reason, we include radiative processes (bremsstrahlung, double
Compton scattering, and three photon annihilation with their inverse
processes) that change the
particle number even though their cross-sections are at least
$\sim\alpha^{-1}\sim 10^2$ times smaller than those of the two-body
processes ($\alpha=e^2/\hbar c=1/137$ is the fine structure
constant). Radiative pair production, $\gamma\gamma\leftrightarrow
e^+e^-\gamma$, which is a radiative variant of photon-photon pair
production, $\gamma\gamma\rightarrow e^+e^-$, is not essential and is
ignored in our simulations.
The processes we include are listed in
Table~1.

\section{The computational method}
In the numerical scheme we define a grid in the $\{{\bf r},
\mu,\epsilon\}$ phase-space as follows. The $r$ domain
($R<r<r_{\rm ext}$) is divided into $j_{\rm max}$ spherical shells
whose boundaries are designated with half integer indices. The $j$
shell ($1\leq j\leq j_{\rm max}$) is between $r_{j-1/2}$ and
$r_{j+1/2}$, with $\Delta r_j=r_{j+1/2}-r_{j-1/2}$ ($r_{1/2}=R$
and $r_{j_{\rm max}+1/2}=r_{\rm ext}$).

The $\mu$-grid is made of $k_{\rm max}$ intervals
$\Delta
\mu_k=\mu_{k+1/2}-\mu_{k-1/2}$: $1\leq k\leq k_{\rm max}$.

The energy grids for photons and
electrons are different. They are
both made of $\omega_{\rm
max}$ energy
intervals $\Delta \epsilon_ {i,\omega}=\epsilon_{i,\omega
+1/2}-\epsilon_{i,\omega-1/2}$: $1\leq\omega \leq \omega_{\rm
max}$,
but the lowest energy for photons is 0, while that for pairs is
$m_ec^2$.

The quantities we compute are the energy densities averaged over
phase-space cells
\begin{equation}
    E_{i,\omega,k,j}(t)
   =\frac{1}{\Delta X}
    \int_{\Delta\epsilon_\omega,\Delta\mu_k,\Delta r_j}
    E_i d\epsilon d\mu r^2dr.
\end{equation}
where $\Delta X=\Delta \epsilon_\omega\Delta \mu_k \Delta (r^3_j)/3$
and
$\Delta (r^3_j)=r^3_{j+1/2}-r^3_{j-1/2}$.

Replacing the space and angle derivatives in the Boltzmann
equations (6) by finite differences, we have the following set of
ordinary differential equations (ODE) for
$E_{i,\omega,k,j}$ specified on the computational grid:
\begin{eqnarray}
  \frac{1}{c}\frac{d E_{i,\omega,k,j}}{d t}
 +\beta_{i,\omega}\frac{\Delta [r^2\mu_k E_{i,\omega,k}]_j}
 {\Delta (r_j^3)/3}
  \nonumber \\
 +\left<\frac{1}{r}\right>_j
  \beta_{i,\omega}\frac{\Delta\left[(1-\mu^2) E_{i,\omega,j}\right]_k}
  {\Delta\mu_k}
  \nonumber \\
 =\sum_q[\eta_{i,\omega,k,j}^q-(\chi E)_{i,\omega,k,j}^q],
 \label{ODE}
\end{eqnarray}
where
$\beta_{\gamma,\omega}=1$,
\begin{equation}
\beta_{e,\omega}
  =\sqrt{1- (m_ec^2/\epsilon_{i,\omega})^2}\,,
\end{equation}
\begin{equation}
\epsilon_{i,\omega}
=(\epsilon_{i,\omega-1/2}+\epsilon_{i,\omega+1/2})/2\,,
\end{equation}
\begin{equation}
\mu_k=(\mu_{k+1/2}+\mu_{k-1/2})/2\,,
\end{equation}
\begin{equation}
\left<\frac{\displaystyle 1}{\displaystyle r}\right>_j
=\frac{\displaystyle (r_{j+1/2}^2-r_{j-1/2}^2)/2}{\displaystyle
(r_{j+1/2}^3-r_{j-1/2}^3)/3}\,,
\end{equation}
\begin{eqnarray}
   \Delta [r^2\mu_k E_{i,\omega,k}]_j
  = r_{j+1/2}^2[\mu_k E_{i,\omega,k}]_{r=r_{j+1/2}}\nonumber\\
   -r_{j-1/2}^2[\mu_k E_{i,\omega,k}]_{r=r_{j-1/2}},
\end{eqnarray}
\begin{eqnarray}
    \Delta\left[(1-\mu^2) E_{i,\omega,j}\right]_k
  = (1-\mu_{k+1/2}^2) [E_{i,\omega,j}]_{\mu=\mu_{k+1/2}}\nonumber\\
   -(1-\mu_{k-1/2}^2) [E_{i,\omega,j}]_{\mu=\mu_{k-1/2}}\,,
\end{eqnarray}
\begin{equation}
    E_{i,\omega,k}(r)
   =\frac{1}{\Delta\epsilon_\omega\Delta\mu_k }
    \int_{\Delta\epsilon_\omega,\Delta\mu_k}
    E_i (\epsilon,\mu ,r) d\epsilon d\mu\,,
\end{equation}
\begin{equation}
    E_{i,\omega,j}(\mu )
   =\frac{3}{\Delta\epsilon_\omega \Delta r_j^3}
    \int_{\Delta\epsilon_\omega,\Delta r_j}
    E_i (\epsilon, \mu ,r) d\epsilon r^2dr\,,
\end{equation}
\begin{eqnarray}
   [\mu_k E_{i,\omega,k}]_{r=r_{j+1/2}}
 = (1-\tilde\chi_{i,\omega,k,j+1/2})
   \nonumber \\
   \times
   \biggl( \frac{\displaystyle \mu_k+|\mu_k|}{\displaystyle 2}
           E_{i,\omega,k,j}
          +\frac{\displaystyle \mu_k-|\mu_k|}{\displaystyle 2}
           E_{i,\omega,k,j+1}
   \biggr)
   \nonumber \\
  +\tilde\chi_{i,\omega,k,j+1/2}\mu_k
   \frac{\displaystyle E_{i,\omega,k,j}+E_{i,\omega,k,j+1}}
   {\displaystyle 2}\,,
\end{eqnarray}

\begin{eqnarray}
\tilde\chi_{i,\omega,k,j+1/2}^{-1}=
  1
 +\frac{1}{\chi_{i,\omega,k,j}\Delta r_j}
 +\frac{1}{\chi_{i,\omega,k,j}\Delta r_{j+1}},
\end{eqnarray}
\begin{eqnarray}
  [ E_{i,\omega,j}]_{\mu=\mu_{k+1/2}}
 = E_{i,\omega,k,j} \nonumber \\
  +\frac{\Delta\mu_k(E_{i,\omega,k,j}-E_{i,\omega,k-1,j})}
   {\Delta\mu_{k-1}+\Delta\mu_k}\,,
\end{eqnarray}
\begin{equation}
\eta_{i,\omega,k,j}^q
  =\frac{\displaystyle 1}
   {\displaystyle \Delta X}
   \int_{\Delta\epsilon_\omega, \Delta\mu_k, \Delta r_j}
   \eta_i^q d\epsilon d\mu
r^2 dr,
\end{equation}
\begin{equation}
(\chi E)_{i,\omega,k,j}^q
  =\frac{\displaystyle 1}
   {\displaystyle \Delta X}
   \int_{\Delta\epsilon_\omega,\Delta\mu_k,\Delta r_j}
   \chi_i E_i^q d\epsilon d\mu
r^2 dr,
\end{equation}
For the physical processes included in our simulations the
expressions $\eta_{i,\omega,k,j}^q$ and $(\chi E)_{i,\omega,k,j}^q$
are given in the Appendix.

The dimensionless coefficient $\tilde\chi$ is introduced
to describe correctly both the optically thin and optically thick
computational cells by means of a compromise between the high order
method and the monotonic transport scheme (see Richtmeyer \& Morton
1967; Mezzacappa \&
Bruenn 1993; Aksenov
1998).

Energy conservation on the grid may be written as
\begin{equation}
    \sum_{i,\omega,k}
    \frac{1}{c}\frac{dE_{i,\omega,k,j}}{d t}
    \Delta\epsilon_\omega\Delta\mu_k
    \frac{\Delta (r_j^3)}{3}
   +F_{j+1/2}-F_{j-1/2}
   =0,
\end{equation}
where $$
  F_{j+1/2}=
  r_{j+1/2}^2\sum_{i,\omega,k}\Delta\epsilon_\omega\Delta\mu_k
      \beta_{i,\omega}[\mu_k E_{i,\omega,k}]_{r=r_{j+1/2}}
$$ is the energy flux through the sphere of radius $r_{j+1/2}$.
In our scheme this is exactly satisfied. Particle number is also
explicitly conserved when only two-body processes are taken into
account; particle production occurs only via three-body processes.

There are several characteristic times in the system. Some are
related to particle reaction times, some to the time to reach
steady
state. These times may greatly differ
from each
other, especially at high luminosities ($\gtrsim 10^{38}$
ergs~s$^{-1}$), when the pair wind is optically thick. The set of
equations (\ref{ODE}), which describes the optically thick wind is
said to be stiff. (A set of differential equations is called stiff
if at least some eigenvalues of
Jacobi matrix differ significantly,
and the real parts of eigenvalues are negative.)
In contrast to Mezzacappa \& Bruenn (1993) we use the Gear's
method (Hall \& Watt 1976) to solve the Boltzmann equations (\ref{ODE}).
This
high order implicit method was developed especially
to find a
solution of stiff sets of ODE. To solve a system of linear
algebraic equations at any time step of the Gear's method
we use the cyclic reduction method (Mezzacappa \& Bruenn 1993).
The number of operations per time step is $\propto
(\omega_\mathrm{max}k_\mathrm{max})^3 j_\mathrm{max}$, which
increases rapidly with the increase of $\omega_{\rm max}$ and
$k_{\rm max}$. Therefore, the numbers of energy and angle
intervals are rather limited in our simulations.

Below, we use the $\epsilon,\mu ,r$-grid with $\omega_\mathrm{max}=
k_\mathrm{max}=8$ and $j_{\rm max}=100$.
The discrete energies of the $\epsilon$-grid are
\begin{equation}
   \epsilon_{i,\omega -1/2}
  = m_ic^2
   +\frac{m_ec^2}{2}\times
    \cases{ 1-\cos\pi x_\omega, & $x_\omega\leq1$, \cr
            3+\cos\pi x_\omega, & $x_\omega>1$,
    }
\end{equation}
and $\epsilon_{i,\omega_{\rm max} +1/2}=\infty$, where $x_\omega=
(\omega -1)/(\omega_\mathrm{max}-3)$. This gives a denser grid at
low energies and near the threshold of pair production,
$\epsilon=m_ec^2$. For photons $(m_\gamma=0)$ the discrete
energies (in keV) of the $\epsilon$-grid  are 0, 48.8, 176.5,
334.4, 462.1, 510.9, 559.7, 687.4, and $\infty$, i.e., three
energy intervals are above the pair production threshold. The
$\mu$-grid is uniform, $\Delta\mu_k=2/k_\mathrm{max}=1/4$. The
shell thicknesses are geometrically spaced, $\Delta r_j=1.3\Delta
r_{j-1}$, and the thickness of the initial shell is $\Delta
r_1=2\times 10^{-4}\mbox{ cm}$.

We ran two test problems. In the first we assumed that the
external boundary is a perfect mirror for both $e^\pm$ pairs and
photons. Pair injection was stopped at some moment, and the
subsequent evolution of the system was followed. We found that it
evolved eventually to the state when the distributions of photons
and pairs are completely isotropic and uniform. In the second we
verified that the processes we included lead to thermal
equilibrium in a spatially uniform system (this also checks the
adequacy of our energy grid). Following Pilla \& Shaham (1997),
we considered the time evolution of photons and pairs that start
far from thermal equilibrium but have isotropic and uniform
distributions. At $t=0$ the total energy density of photons and
pairs was taken to be that in thermal equilibrium at $T=10^9$ K
($k_{\rm B}T\simeq 100$ keV).
The initial photon and the pair distributions were flat and 
nonzero within the the energy interval $10^{-2}\leq
(\epsilon_i-m_ic^2)/m_ec^2\leq 2$. The initial densities of
photons and pairs were equal.
As seen in Figures 2 and 3, at time $t=2.7\times 10^{-8}$~s, when
the system is almost stationary, photons and pairs are near
thermal equilibrium. For photons and electrons, the differences
between the calculated and equilibrium spectral energy densities
are a few tens percents (attributed to the coarseness of the grid)
which is then the accuracy of calculations of energy spectra in
our simulations.

To check the effects of grid coarseness we also performed test
computations with $k_{\rm max}=4$ and 6, and separately with
$\omega_{\rm max}=6$ and 7 for the discrete energies of the
$\epsilon$-grid given by equation (30) where $x_\omega =(\omega -1)/
(\omega_{\rm max}-2)$, i.e., only two high-energy intervals are
above the pair production threshold. We did not observe
essential changes in the results. This implies that
the main results of our simulations are not sensitive to the
$\epsilon,\mu$-grid in spite of its being rather coarse.

\section{Numerical results}
We give here the results for the properties of  pair winds. The
energy injection rate, $\dot {\rm E}$, is the only parameter we
vary in our simulations. As explained in \S~2, we start from an
empty wind injecting pairs at a rate $10^{34}$ ergs~s$^{-1}$.
After a steady state is reached we start a new run with this
steady state as initial condition, increase the energy injection
rate by a factor of 10, wait for steady state, and so on. Figure~4
shows the total emerging luminosity in photons and in pairs at the
external boundary as a function of time $t$. We see that in each
run this luminosity increases eventually to its maximum value $L$,
which equals the energy injection rate, $L=L_e+L_\gamma=\dot{\rm
E}$. The rise time is $\sim 10^{-3}-10^{-2}$ s, which is the
characteristic time on which the pair wind becomes stationary in
the examined space domain. We next present the results for the
structure of the stationary winds and their emerging emission.
Figure~5 shows the mean optical depth for photons, from $r$ to
$r_{\rm ext}$, defined as
\begin{equation}
\tau_\gamma (r) = \int_r^{r_{\rm ext}}dr \left[{\int d\epsilon\,
d\mu \,\chi_\gamma\, E_\gamma (\epsilon,\mu,r) \over \int
d\epsilon\, d\mu\, E_\gamma (\epsilon,\mu,r)}\right],
\end{equation}
where
\begin{equation}
\chi_\gamma=\sum_q \chi_\gamma^q\,.
\end{equation}
The contribution from $r_{\rm ext}$ to infinity is negligible for
$r < 10^8$ cm, so $\tau_\gamma (r)$ is practically the mean
optical depth from $r$ to infinity for these values of $r$. The
pair wind is optically thick [$\tau_\gamma (R)>1$] for $\dot {\rm
E}>10^{37}$ ergs~s$^{-1}$. The radius of the wind photosphere
$r_{\rm ph}$, determined by condition $\tau (r_{\rm ph})=1$,
varies from $\sim R$ for $\dot{\rm E}=10^{37}$ ergs~s$^{-1}$ to
$\sim 10R\simeq 10^7$ cm for $\dot{\rm E}=10^{42}$
ergs~s$^{-1}$. The wind photosphere is always deep inside our
chosen external boundary ($r_{\rm ph} < 0.1r_{\rm ext}$),
justifying our neglect of the inward $(\mu <0)$ fluxes at
$r=r_{\rm ext}$.

The emerging luminosities in $e^\pm$ pairs ($L_e$) and photons
($L_\gamma$) are shown as fractions of the total luminosity
$L=L_e+L_\gamma$ in Figure~6. For $L < L_{\rm eq}\simeq
2\times 10^{35}$ ergs~s$^{-1}$, the injected pairs remain
intact, and they dominate in the emerging emission ($L_e>L_\gamma$).
For $L
>L_{\rm eq}$, the emerging
emission consists mostly of photons ($L_\gamma >L_e$). This simply
reflects the fact that in this case the pair annihilation time
$t_{\rm ann}\sim (n_e \sigma_{\rm T}c)^{-1}$ is less than the
escape time $t_{\rm esc}\sim R/c$, so most injected pairs
annihilate before they escape ($\sigma_{\rm T}$ is the Thomson
cross section). The condition $t_{\rm ann}\simeq t_{\rm esc}$
implies (e.g., Beloborodov 1999 and references therein)
\begin{equation}
L\simeq {2\pi m_ec^3R\over \sigma_{\rm T}}\simeq 2.3\times
10^{35}\,\,{\rm ergs~s}^{-1}\,,
\end{equation}
which practically coincides with $L_{\rm eq}$.
For very high luminosities $(L\gg L_{\rm eq})$,
reconversion of photons into pairs is inefficient, as the mean photon
energy at the wind photosphere is rather below the pair-creation threshold
(see below), and photons strongly dominate in the emerging emission,
$L_\gamma\simeq \dot {\rm E}\gg L_e$.

Figure~7 shows the rates of outflow of $e^\pm$ pairs ($\dot N_e$)
and photons ($\dot N_\gamma$) through the surface at radius $r$ as
functions of $r$. At $r=R$, the pair outflow rate is put equal to
the rate of pair injection
\begin{equation}
\dot N_e^{\rm in}\simeq10^{48}(\dot {\rm E}/ 10^{42}
\,{\rm ergs~s}^{-1})\,\, {\rm s}^{-1},
\end{equation}
while the photon outflow rate is zero ($\dot N_\gamma^ {\rm
in}=0$). (Because the surface temperature depends weakly on $\dot
{\rm E}$, $\dot N_e^{\rm in}$ is nearly proportional to $\dot {\rm
E}$.) There is the upper limit on the rate of emerging pairs $\dot
N_e^{\rm max}\simeq 10^{43}$ s$^{-1}$. If $\dot N_e^{\rm in}\gg
\dot N_e^{\rm max}$, the rate of pair outflow $\dot N_e$ decreases
sharply at the distance
\begin{equation}
l_{\rm ann}= {1\over n_{e,s}\sigma_{\rm
ann}}\simeq 10^{-2}\left( {\dot{\rm E}\over 10^{42}\,{\rm
ergs~s}^{-1}}\right)^{-1} \,{\rm cm}
\end{equation}
from the stellar surface, where $n_{e,s}= \dot N^{\rm in}_e/4\pi
R^2v^{\rm out}_{e,s}$ is the density of pairs at the surface,
$v^{\rm out}_{e,s}\simeq 10^9$ cm~s$^{-1}$ is the velocity of the
pair plasma outflow near the surface (see below), $\sigma_{\rm
ann}\simeq \sigma_{\rm T} (c/\langle v_e \rangle)$ is the pair
annihilation cross section,
and $\langle v_e \rangle \simeq (3k_{\rm B} T_{\rm
S}/m_e)^{1/2}\simeq 10^{10}$ cm~s$^{-1}$ is the mean velocity of
injected pairs. As a result, the rate of emerging pairs is limited
to $\dot N_e^{\rm max}$ within a factor of 2. For $\dot{\rm
E}\gtrsim 10^{38}$ ergs~s$^{-1}$, radiative three-body processes
are important, and the total rate of the particle outflow $\dot
N=\dot N_e+\dot N_\gamma$ increases with radius (see Fig.~7). For
$\dot {\rm E}=10^{42}$ ergs~s$^{-1}$, when photons strongly
prevail in the emerging emission, the rate of emerging photons
increases by a factor of 15 in comparison with $\dot N_e^{\rm
in}$. The rates of energy outflow in $e^\pm$ pairs $(\dot {\rm
E}_e)$ and photons $(\dot{\rm E}_\gamma)$ vary with radius more or
less similarly to the particle outflow rates, except that the
total energy rate $(\dot {\rm E}_e+\dot{\rm E}_\gamma)$ doesn't
depend on radius and equals to the energy injection rate $\dot{\rm
E}$ (see Fig.~8). The upper limit on the emerging luminosity in
$e^\pm$ pairs is $L_e^{\rm max}\simeq 2\times 10^{37}$
ergs~s$^{-1}$ (which includes the rest energies). The existence of
the upper limits $\dot N_e^{\rm max}$ and $L_e^{\rm max}$ are
connected with production of photons via radiative three-body
processes. This can be seen in Figure~9, which shows the rates of
particle outflow when only two-body processes are included. We see
that for $\dot{\rm E}\gg 10^{37}$ ergs~s$^{-1}$ and $r-R\gg l_{\rm
ann}$ the outflow rate of $e^\pm$ pairs is only a few times less
than that of photons.

The bulk velocity of the pair plasma outflow
\begin{equation}
v^{\rm out}_e =\dot N_e/4\pi r^2n_e
\end{equation}
is shown in Figure~10, where $n_e$ is the pair number density. We
see that at low luminosities ($\dot {\rm E}\lesssim 10^{37}$
ergs~s$^{-1}$) $v^{\rm out}_e$ increases with distance from the
surface, reaching, for $\dot{\rm E}=10^{37}$ ergs~s$^{-1}$,
 as high a value as $0.7c$. For such luminosities the
wind is optically thin, $\tau_\gamma\lesssim 1$ (see Fig.~5), and
pairs and photons flow away more or less independently. The
increase of $v^{\rm out}_e$ occurs because outflowing pairs are
heated by annihilation photons via Compton scattering. At high
luminosities $(\dot {\rm E} \gtrsim 10^{38}$ ergs~s$^{-1}$) the
velocity of the pair plasma outflow decreases at $r-R\gtrsim
l_{\rm ann}$ and is as small as $\sim 2\times 10^7$ cm~s$^{-1}$ for
$\dot
{\rm E}\simeq 10^{41}-10^{42}$ ergs~s$^{-1}$ (see Fig.~10).
To
explain the decrease of $v_e^{\rm out}$ we introduce the bulk
velocity of the photon gas outflow
\begin{equation}
v^{\rm out}_\gamma =\dot N_\gamma/4\pi r^2n_\gamma\,,
\end{equation}
where $n_\gamma$ is the photon number density. The ratio
$v_\gamma^{\rm out}/c$ is a measure of photon anisotropy and
varies from zero at the stellar surface to 1 far from the wind
photosphere (see Fig.~11). In our case when the injected
plasma consists of pairs with mean velocity $\langle
v_e\rangle\sim c$, the free path length of Compton
scattering,
which is the main mechanism of opacity for photons, is of the
order of the free path length of pair annihilation
$l_{\rm ann}$.
Therefore, at the distance of $\sim l_{\rm ann}$ from the stellar
surface the outflowing pair plasma is decelerated by nearly
isotropic ($v_\gamma^{\rm out}/c\ll 1)$ photons. Then, the pair
plasma, which is optically thick, flows away with nearly the same
velocity as the photon gas ($v^{\rm out}_e\simeq v^{\rm
out}_\gamma$) up to the wind photosphere (see Figs.~10 and 11). It
is worth noting that for $\dot{\rm E} \gtrsim 10^{38}$
ergs~s$^{-1}$ and $r-R\gg l_{\rm ann}$, when photons prevail, the
wind dynamics is mostly determined by photons, pairs being only
responsible for photon opacity.

Figure 12 shows the pair number density as a function of the distance
from the stellar surface. As seen in Figures 7, 10 and 12, for
$\dot N_e^{\rm in}\gg \dot N_e^{\rm max}$ the rate of pair number
outflow $\dot N_e=4\pi r^2n_e v_e^{\rm out}$
decreases very fast with radius at $r-R\gtrsim l_{\rm ann}$
because both $n_e$ and $v_e^{\rm out}$ decrease fast at the same
distances.

Figures 13 and 14 show, respectively, the mean energies of photons
and pairs, which are determined by
\begin{equation}
\langle \epsilon_i\rangle ={\int \epsilon_i^3 f_i(\epsilon_i, \mu,
r) \,d\epsilon_i\,d\mu \over \int \epsilon_i^2 f_i(\epsilon_i, \mu, r)
\,d\epsilon_i\,d\mu} -m_ic^2\,,
\end{equation}
as functions of the distance from the surface. We see that the
mean photon energy $\langle\epsilon_\gamma\rangle$ decreases with
the increase of $\dot{\rm E}$ at all radii. In the $\dot {\rm E}$
range from $\sim 10^{34}$  to $\sim 10^{37}$ ergs~s$^{-1}$, where
most of the photons in the system are produced via pair
annihilation, the decrease of $\langle\epsilon_\gamma\rangle$ is
rather weak and occurs because of energy transfer from
annihilation photons to $e^\pm$ pairs via Compton scattering. As a
result, the emerging pairs are heated up to a mean kinetic energy
$\langle \epsilon_e\rangle \simeq 500$ keV at $\dot{\rm E}\simeq
10^{37}$ ergs~s$^{-1}$ (see Fig. 14). For $\dot{\rm E}\gtrsim
10^{38}$ ergs~s$^{-1}$, $\langle\epsilon_\gamma\rangle$ decreases
mainly because of production of rather low-energy photons in the
radiative three-body processes (see Fig. 15). For $\dot{\rm
E}=10^{42}$ ergs~s$^{-1}$ we have
$\langle\epsilon_\gamma\rangle\sim 40$ keV for the emerging
photons. This value of $\langle\epsilon_\gamma \rangle$ is near
the mean energy of blackbody photons for the same energy density
as that of the photons at the wind photosphere, which is
$\langle\epsilon_\gamma^{\rm bb}\rangle\simeq 30$ keV. (The
difference between $\langle\epsilon_\gamma \rangle$ and
$\langle\epsilon_\gamma^{\rm bb}\rangle$ is less than the energy
resolution of the $\epsilon$-grid at low energies, which is $\sim
20$ keV.)

Our Figure~15 shows that for high luminosities ($L> 2\times 10^{37}$
ergs~s$^{-1}$) the mean energies of photons and pairs decrease
with luminosity. This is consistent with previous studies showing
a similar behavior for large enough values of the compactness
parameter
\begin{equation}
l={L\sigma_{\rm T}\over m_ec^3R}=2\pi {L\over L_{\rm eq}}
\end{equation}
(Svensson 1984 and references therein).

Figures 16 and 17 present the energy spectra of the emergent
photons and  pairs. At low luminosities, $\dot{\rm E}\sim
10^{34}-10^{37}$ ergs~s$^{-1}$, photons that form by pair
annihilation escape more or less freely, and the photon spectra
resemble a very wide annihilation line. For $\dot {\rm E}\gtrsim
10^{38}$ ergs~s$^{-1}$, changes in the particle number due to
radiative three-body processes are essential, and their role in
thermalization of the outflowing plasma increases with the
increase of $\dot{\rm E}$. We see in Figure 14 that for $\dot {\rm
E} = 10^{42}$ ergs~s$^{-1}$ the photon spectrum is near blackbody,
except for the presence of a high-energy tail at $\epsilon_\gamma
>100$ keV. Such a hard spectrum of photons together with the
anti-correlation between spectral hardness and photon luminosity
(see Fig.~15) could be a good observational signature of a hot,
bare, strange star.

The high-energy tail of the emergent photons covers six
$\epsilon$-grid intervals and is real. Moreover, if the
energy injection rate $\dot {\rm E}$ is less than $10^{43}$
ergs~s$^{-1}$, such a high-energy tail has to be present, i.e.,
the energy spectrum of the emergent photons cannot be completely
a Planckian at high energies. Indeed, if at the wind
photosphere, photons and pairs are in thermal equilibrium,
their temperature is
\begin{equation}
T_{\rm ph}^{\rm eq}\simeq\left({\dot{\rm E}
\over 4\pi r_{\rm ph}^2\sigma}\right)^{1/4}\,.
\end{equation}
For $\dot{\rm E}=10^{42}$ ergs~s$^{-1}$,
when $r_{\rm ph}$ is $\sim 10^7$ cm (see
Fig.~5), we have $T_{\rm ph}^{\rm eq}\simeq 0.61\times 10^8$ K. At this
temperature the density of equilibrium pairs is (e.g., Paczy\'nski 1986)
\begin{equation}
n_{e,{\rm ph}}\simeq 4.4\times 10^{30}\left({kT_{\rm ph}^{\rm eq}\over
m_ec^2}\right)^{3/2}\exp \left({- m_ec^2\over kT_{\rm ph}^{\rm eq}}
\right),
\end{equation}
or numerically, $n_{e,{\rm ph}}\sim 10^{15}$ cm$^{-3}$,
while for $\dot{\rm E}>
10^{37}$ ergs~s$^{-1}$ the pair density at the photosphere cannot be
essentially smaller than $\sim 1/(r_{\rm ph}\sigma_{\rm ann})\sim
10^{17}
-10^{18}$ cm$^{-3}$ (see Fig.~12), at which the optical depth for pair
annihilation is $\sim 1$. Even if we take the stellar radius as the
radius of the wind photosphere, $r_{\rm ph}=R$, for $\dot {\rm E}=10^{42}$
ergs~s$^{-1}$ from equations (40) and (41) we have $T_{\rm ph}^{\rm eq}
\simeq 1.93\times 10^8$ K and $n_{e,{\rm ph}}\simeq 10^{15}\ll
1/(r_{\rm ph}\sigma_{\rm ann})$. Hence,
in this case the energy spectrum of the emergent photons cannot be
a Planckian at high energies, and it has to range to $\sim m_ec^2$.
This spectrum might be completely a Planckian starting only from
$\dot {\rm E}\simeq 10^{43}$ ergs~s$^{-1}$,
at which for $r\simeq R$,  we have $T_{\rm ph}^{\rm eq}\simeq
3\times 10^8$ K and $n_{e,{\rm ph}}\gtrsim 1/(r_{\rm ph}\sigma_{\rm ann})$.

\section{Discussion}
We have identified certain characteristics of the expected
radiation from  hot, bare, strange stars that, we hope, will help
identify such stars if they exist. The spectrum, we find, is
rather hard for the studied luminosity range. This makes such
stars amenable to detection and study by sensitive, high energy
instruments, such as INTEGRAL (e.g., Schoenfelder 2001), which is
more
sensitive in this range than previous detectors.

As the emission from bare, strange stars is characterized by
super-Eddington luminosities (see Fig.~1) and, as we now find,
also by hard X-ray
spectra, soft
$\gamma$-ray repeaters (SGRs), which are the sources of short
bursts of hard X-rays with super-Eddington luminosities (up to
$\sim 10^{42}-10^{45}~{\rm ergs~s}^{-1}$), are potential
candidates for strange stars (e.g., Alcock, Farhi, \& Olinto
1986b; Cheng \& Dai 1998; Usov 2001b). The bursting activity of
SGRs may be explained by fast heating of the stellar surface up to
the temperature of $\sim (1-2)\times 10^9$~K and its subsequent
thermal emission (Usov 2001b,c). The heating mechanism may be
either
impacts of comets onto bare strange stars (Zhang, Xu, \&
Qiao 2000;
Usov 2001b) or fast decay of superstrong ($\sim 10^{14}-10^{15}$~G)
magnetic fields (Usov 1984; Thompson \& Duncan 1995; Heyl \& Kulkarni
1998).
For typical luminosities of SGRs ($L\sim 10^{41}- 10^{42}$
ergs~s$^{-1}$),  the mean photons energy we find is $\sim 40$ keV
(see Fig.~15), which is consistent with observations of SGRs
(Hurley 2000).
The rise time ($\sim 10^{-3}-10^{-2}$ s) of the
emerging luminosity (see Fig.~4) implies that the variability of
bursts from SGRs may be explained in the strange star model.

Another important idiosyncrasy that we find is a strong
anti-correlation between spectral hardness and luminosity. While
at very high luminosities ($L>10^{42}-10^{43}$ ergs~s$^{-1}$) the
spectral temperature increases with luminosity as in blackbody
radiation, in the range of luminosities we studied, where thermal
equilibrium is not achieved, the expected correlation is opposite
(see Fig.~15). Such anti-correlations were, indeed, observed for
SGR 1806-20 and SGR 1900+14 where the burst statistic is high
enough (e.g., Feroci et al. 2001; Gogus et al. 2001; Ibrahim et
al. 2001). This is encouraging, but a direct comparison of this
data with results such as ours will require a more detailed
analysis. In particular, the effects of strong magnetic fields
have to be included. The processes we consider can be
significantly modified by such fields, especially above $10^{13}$
G (e.g., Daugherty \& Harding 1989). Qualitatively new processes
such as photon splitting (e.g., Adler 1971; Baring \& Harding
2001; Usov 2002) may become important in a strong magnetic field.
The observed periodic variations in the SGRs may be due to a
rotation of a star with non-uniform surface temperature, while our
result apply for isotropic emission. We hope to deal with these
effects elsewhere.

We noted that there is an upper limit to the rate of emerging
pairs $\dot N_e^{\rm max}\simeq 10^{43}$~s$^{-1}$. Positrons can
then annihilate in an ambient medium. Hence, the luminosity in
annihilation emission from the region far from a hot bare strange
star may be as high as $2mc^2\dot N_e^{\rm max}\simeq 10^{37}$
ergs~s$^{-1}$.

In our simulations, gravity is neglected. Gravity  decelerates the
pair wind and increases the pair density near the surface. Also,
photons are red shifted when they escape from the star's vicinity.
To estimate the effects of gravity we ran simulations for a star
mass of $1.4M_\odot$, where gravity was included in the Newtonian
approximation. We found that at low luminosities ($\dot {\rm E}
\lesssim \tilde L_{\rm Edd}\simeq 10^{35}$ ergs~s$^{-1}$) the pair
 density $n_e$ does indeed increase significantly near the
surface, and for $\dot {\rm E}=10^{34}$ ergs~s$^{-1}$ the value of
$n_e$ is about ten times higher than without gravity. At such low
luminosities the velocity of the pair plasma outflow deceases by
about a factor of 3 because of gravity effects and is equal to
$10^9$ cm~s$^{-1}$ within a factor of 2 (cf. Fig.~10). The
probability of pair annihilation increases because of the pair
density increase, and photons dominate the emerging emission for
$L=\dot{\rm E}>3\times 10^{34}$ ergs~s$^{-1}$ (cf. Fig.~6). At
high luminosities ($\dot{\rm E}>\tilde L_{\rm Edd}$) gravity
doesn't affect the pair wind structure significantly, especially
at $\dot{\rm E}\gg L_{\rm Edd}$. For $\dot{\rm E}< 10^{37}$
ergs~s$^{-1}$ when the pair wind is optically thin (see Fig.~5)
the mean energy of emergent photons
$\langle\epsilon_\gamma\rangle$ deceases by $\sim 20$\% because of
the red shift, while for $\dot{\rm E}> 10^{38}$ ergs~s$^{-1}$ the
decrease of $\langle\epsilon_\gamma\rangle$ is less than a few
percents, compared with the accuracy of our simulations, which is
not higher than $\sim 10-20$\% because the $\epsilon,\mu$-grid is
rather coarse. Note also that gravity corrections in the Newtonian
approximation are valid for the relativistic Boltzmann equations
with an accuracy of a factor of two, and, while capturing some of
the effects of gravity, they are not completely self-consistent.
We have thus preferred to present our results for the self
consistent case without gravity, deferring the full inclusion of
relativistic gravity for a later treatment.

\begin{acknowledgements}
We are grateful to the anonymous referee for a careful reading of
the manuscript and for many helpful comments.
This work was supported by the Israel Science Foundation of
the Israel Academy of Sciences and Humanities.
\end{acknowledgements}

\begin{appendix}
\centerline{APPENDIX}
\section{Emission and absorption coefficients
for two- and three-body processes}
We use two sets of independent variables ${\bf p}, r,t$, and
$\epsilon,\mu,r,t$. We take a uniform distribution of particle
density inside the grid volume
$\Delta\epsilon_{\nu,\omega}\Delta\mu_k\Delta (r_j^3)/ 3$:
\begin{equation}
n_i (\epsilon,\mu,r)=
{E_i(\epsilon,\mu,r)\over\epsilon}
   =\frac{E_{i,\omega,k,j}}{\epsilon_{i,\omega}}\,\,\,
    \mbox{ for }
    \epsilon_{i,\omega-1/2}<\epsilon<\epsilon_{i,\omega+1/2},
    \mbox{ }\,\,
    \mu_{k-1/2}<\mu<\mu_{k+1/2},
    \mbox{ }\,\,
    r_{j-1/2}< r< r_{j+1/2}, \label{Edistribution}
\end{equation} where
$\epsilon_{i,\omega}=\epsilon_{i,\omega-1/2}+\Delta\epsilon_{i,\omega}/2$.
(We suppress the $r$-dependence of functions.)

\subsection{Compton scattering}
The time evolution of the distribution functions of photons and
pair
particles due to Compton scattering, $\gamma e\rightarrow \gamma' e'$,
may be described by
(Ochelkov et al 1979; Berestetskii et al. 1982)
\begin{equation}
  \left(\frac{\partial f_\gamma(\mathbf{k},t)}
  {\partial t}\right)_\mathrm{\gamma e\rightarrow\gamma'e'}
 =\int d\mathbf{k}'d\mathbf{p}d\mathbf{p}'
  w_{\mathbf{k}',\mathbf{p}';\mathbf{k},\mathbf{p}}
  [ f_\gamma({\mathbf{k}',t)}f_e({\mathbf{p}'},t)
        -f_\gamma(\mathbf{k},t)f_e(\mathbf{p},t)], \label{Comptongamma}
\end{equation}
\begin{equation}
  \left(\frac{\partial f_e(\mathbf{p},t)}
  {\partial t}\right)_\mathrm{\gamma e\rightarrow\gamma'e'}
 =\int d\mathbf{k}d\mathbf{k}'d\mathbf{p}'
  w_{\mathbf{k}',\mathbf{p}';\mathbf{k},\mathbf{p}}
  [f_\gamma({\mathbf{k}'},t)f_e({\mathbf{p}'},t)
   -f_\gamma(\mathbf{k},t)f_e(\mathbf{p},t)], \label{Comptone}
\end{equation}
where
\begin{equation}
  w_{\mathbf{k}',\mathbf{p}';\mathbf{k},\mathbf{p}}
  =\frac{c\delta(\epsilon_\gamma-\epsilon_e-\epsilon_\gamma'-\epsilon_e')}
   {(2\pi\hbar)^2}
   \delta(\mathbf{k}+\mathbf{p}-\mathbf{k}'-\mathbf{p}')
   \frac{|M_{fi}|^2}
   {16\epsilon_\gamma\epsilon_e\epsilon_\gamma'\epsilon_e'},
   \label{w}
\end{equation}
is the probability of the process,

\begin{eqnarray}
  |M_{fi}|^2
      =2^7(\pi\hbar)^2r_e^2m_e^2c^7
             \Biggl\{ \frac{m_e^2c^2}{s-m_e^2c^2}
                     +\frac{m_e^2c^2}{u-m_e^2c^2}
                     +\left( \frac{m_e^2c^2}{s-m_e^2c^2}
                            +\frac{m_e^2c^2}{u-m_e^2c^2}
                      \right)^2 \nonumber \\
                     -\frac{1}{4}
                      \left( \frac{s-m_e^2c^2}{u-m_e^2c^2}
                            +\frac{u-m_e^2c^2}{s-m_e^2c^2}
                      \right)
             \Biggr\}, \label{M_fi_gamma}
\end{eqnarray}
is the square of the matrix element $M_{fi}$,
$r_e=\displaystyle e^2/(m_e c^2)$ is the classical electron radius,
$s=(\mathfrak{p}+\mathfrak{k})^2$ and
$u=(\mathfrak{p}-\mathfrak{k}')^2$ are invariants,
$\mathfrak{k}=(\epsilon_\gamma/c)(1,\mathbf{e}_\gamma)$ and
$\mathfrak{p}=(\epsilon_e/c)(1,\beta\mathbf{e}_e)$ are
energy-momentum four vectors of photons and electrons, respectively,
$d\mathbf{p}=d\epsilon do\epsilon^2\beta_\nu/c^3$, and $do=d\mu
d\phi$.

As an example, we calculate the value $(\chi E)_{\gamma,
\omega,k}^\mathrm{\gamma e\rightarrow\gamma'e'}$ . We start this
calculation by considering the negative term in equation
(\ref{Comptongamma}), which is responsible for the Compton
absorption of photons:
\begin{equation}
  \left(\frac{\partial f_\gamma(\mathbf{k},t)}
  {\partial t}\right)_\mathrm{\gamma
  e\rightarrow\gamma'e'}^\mathrm{abs}=
  -\int d\mathbf{k}'d\mathbf{p}d\mathbf{p}'
  w_{\mathbf{k}',\mathbf{p}';\mathbf{k},\mathbf{p}}
  f_\gamma(\mathbf{k},t)f_e(\mathbf{p},t)\,.\label{df-abs}
  \end{equation}
  Substituting equation (\ref{w}) into equation (\ref{df-abs}) we can
  obtain
  \begin{equation}
  \left(\frac{\partial f_\gamma(\mathbf{k},t)}
  {\partial t}\right)_\mathrm{\gamma
  e\rightarrow\gamma'e'}^\mathrm{abs}
 =-\int do_\gamma'd\mathbf{p}
  \frac{c\epsilon_\gamma' |\tilde M_{fi}|^2}
  {16\epsilon_e\epsilon_\gamma
   \epsilon_e'}
  f_\gamma(\mathbf{k},t)f_e(\mathbf{p},t),
\end{equation}
where
\begin{equation}
\epsilon_\gamma'
 =\frac{\displaystyle \epsilon_e\epsilon_\gamma(1-
 \beta\mathbf{b}_e\mathbf{\cdot}\mathbf{b}_\gamma)}
  {\displaystyle  \epsilon_e(1-\beta\mathbf{b}_e\mathbf{\cdot}
  \mathbf{b}_\gamma')
   +\epsilon_\gamma(1-\mathbf{b}_\gamma\mathbf{\cdot}\mathbf{b}_\gamma')
  }\,,\,\,\,\,\,
\epsilon_e'
 =\epsilon_e+\epsilon_\gamma-\epsilon_\gamma'\,,
\end{equation}
$\mathbf{b}_i=\mathbf{p}_i/p$, $\mathbf{b}_i'=\mathbf{p}_i'/p'$,
$\mathbf{b}_e'
 =(\beta\epsilon_e\mathbf{b}_e+\epsilon_\gamma\mathbf{b}_\gamma-
 \epsilon_\gamma'\mathbf{b}_\gamma')
 /(\beta'\epsilon_e')$,
 and $|\tilde
M_{fi}|^2=|M_{fi}|^2/[c^3(2\pi\hbar)^2]$.

For photons, the absorption coefficient in the Boltzmann equations
(1) is
\begin{eqnarray}
  \chi_{\gamma}^\mathrm{\gamma e\rightarrow\gamma'e'}f_\gamma
 =-\frac{1}{c}\left(\frac{\partial f_\gamma}{\partial
  t}\right)_\mathrm{\gamma
  e\rightarrow\gamma'e'}^\mathrm{abs}
 =\int
  dn_e
  do_\gamma'
   \frac{\epsilon_\gamma'|\tilde M_{fi}|^2}
   {16\epsilon_e\epsilon_\gamma
    \epsilon_e'}f_\gamma,
\label{chi-fgamma}
\end{eqnarray}
where $dn_i=d\epsilon_i do_i\epsilon_i^2\beta_i
f_i/c^3=d\epsilon_i do_i E_i/(2\pi\epsilon_i)$.

From equations (\ref{Comptongamma}) and (\ref{chi-fgamma}), we can
write the
absorption coefficient for photon energy density $E_\gamma$
averaged over the $\epsilon,\mu$-grid with zone numbers $\omega$
and $k$ as
\begin{eqnarray}
  (\chi E)_{\gamma,\omega,k}^\mathrm{\gamma e\rightarrow\gamma'e'}
 \equiv\frac{
 \int_{\epsilon_\gamma\in\Delta\epsilon_{\gamma,\omega}\atop\mu_
 \gamma\in\Delta\mu_k}
 d\epsilon_\gamma d\mu_\gamma
 (\chi E)_\gamma^\mathrm{\gamma e\rightarrow\gamma'e'}}
  {\Delta\epsilon_{\gamma,\omega}\Delta\mu_k}
 =\frac{1}{\Delta\epsilon_{\gamma,\omega}\Delta\mu_k}
  \int_{\epsilon_\gamma\in\Delta\epsilon_{\gamma,\omega}
        \atop
        \mu_\gamma\in\Delta\mu_{k}}
  dn_\gamma dn_e do_\gamma'
   \frac{\epsilon_\gamma'|\tilde M_{fi}|^2}
   {16\epsilon_e
    \epsilon_e'}.\label{chiE1}
\end{eqnarray}
Similar integrations can be performed for the other terms of equations
(\ref{Comptongamma}), (\ref{Comptone}), and we have
\begin{eqnarray}
  \eta_{\gamma,\omega,k}^\mathrm{\gamma e\rightarrow\gamma'e'}
 \equiv\frac{\int_{\epsilon_\gamma\in\Delta\epsilon_{\gamma,\omega}
 \atop\mu_\gamma\in\Delta\mu_k}
 d\epsilon_\gamma d\mu_\gamma
 \eta_{\gamma}^\mathrm{\gamma e\rightarrow\gamma'e'}}
  {\Delta\epsilon_{\gamma,\omega}\Delta\mu_k}
 =\frac{1}{\Delta\epsilon_{\gamma,\omega}\Delta\mu_k}
  \int_{\epsilon_\gamma'\in\Delta\epsilon_{\gamma,\omega}
        \atop
        \mu_\gamma'\in\Delta\mu_{k}}
  dn_\gamma dn_e do_\gamma'
   \frac{\epsilon_\gamma'^2|\tilde M_{fi}|^2}
   {16\epsilon_e\epsilon_\gamma
    \epsilon_e'},
   \label{C_eta_gamma}
\end{eqnarray}
\begin{eqnarray}
  \eta_{e,\omega,k}^\mathrm{\gamma e\rightarrow\gamma'e'}
 \equiv\frac{\int_{\epsilon_e\in\Delta\epsilon_{e,\omega}\atop\mu_e
 \in\Delta\mu_k}
 d\epsilon_e d\mu_e
 \eta_{e}^\mathrm{\gamma e\rightarrow\gamma'e'}}
 {\Delta\epsilon_{e,\omega}\Delta\mu_k}
 =\frac{1}{\Delta\epsilon_{e,\omega}\Delta\mu_k}
  \int_{\epsilon_e'\in\Delta\epsilon_{e,\omega}
        \atop
        \mu_e'\in\Delta\mu_{k}}
  dn_\gamma dn_e do_\gamma'
   \frac{\epsilon_\gamma'|\tilde M_{fi}|^2}
   {16\epsilon_e\epsilon_\gamma},
   \label{C_eta_e}
\end{eqnarray}
\begin{eqnarray}
  (\chi E)_{e,\omega,k}^\mathrm{\gamma e\rightarrow\gamma'e'}
 \equiv\frac{\int_{\epsilon_e\in\Delta
 \epsilon_{e,\omega}\atop\mu_e\in\Delta\mu_k}
 d\epsilon_e d\mu_e
  (\chi E)_e^\mathrm{\gamma e\rightarrow\gamma'e'}}
  {\Delta\epsilon_{e,\omega}\Delta\mu_k}
 =\frac{1}{\Delta\epsilon_{e,\omega}\Delta\mu_k}
  \int_{\epsilon_e\in\Delta\epsilon_{e,\omega}
        \atop
        \mu_e\in\Delta\mu_{k}}
  dn_\gamma dn_e do_\gamma'
   \frac{\epsilon_\gamma'|\tilde M_{fi}|^2}
   {16\epsilon_\gamma
    \epsilon_e'}.\label{chiE2}
\end{eqnarray}
The emission and absorption coefficients in the Boltzmann equations
(12) are given by
\begin{equation}
\eta^q_{i,\omega,k,j}={1\over \Delta (r^3_j)/3}
\int_{r\in\Delta r_j}\eta^q_{i,\omega,k}r^2dr\,,\,\,\,\,\,\,\,
(\chi E)^q_{i,\omega,k,j}={1\over \Delta (r^3_j)/3}
\int_{r\in\Delta r_j}(\chi E)^q_{i,\omega,k}r^2dr\,,
\end{equation}
where $q$ is $\gamma e\rightarrow \gamma'e'$ for Compton scattering.

To integrate equations (\ref{chiE1})-(\ref{chiE2}) numerically
over $\phi$ ($0\leq \phi\leq 2\pi$) we introduce a uniform grid
$\phi_{l\mp1/2}$ with $1\leq l\leq l_\mathrm{max}$ and $\Delta
\phi_l=(\phi_{l+1/2}-\phi_{l-1/2})/2=2\pi/l_{\rm max}$. We assume
that any function of $\phi$ in equations
(\ref{chiE1})-(\ref{chiE2}) in the interval $\Delta \phi_j$ is
equal to its value at $\phi=\phi_j
=(\phi_{l-1/2}+\phi_{l+1/2})/2$. Since the problem is
axi-symmetric it is necessary to integrate over $\phi$ only once
at the start of calculations. The number of intervals of the
$\phi$-grid is taken as $l_\mathrm{max}=2k_{\rm max}=16$.

\subsection{Two-photon pair annihilation and creation}
 The rates of change of the distribution function due to
  $e^-e^+\rightleftarrows\gamma_1\gamma_2$
  are
\begin{equation}
  \left(\frac{\partial f_{\gamma_i}(\mathbf{k}_i,t)}
  {\partial t}\right)_{e^-e^+\rightarrow\gamma_1\gamma_2}
 = \int d\mathbf{k}_j d\mathbf{p}_-d\mathbf{p}_+
   w_{\mathbf{k}_1,\mathbf{k}_2;\mathbf{p}_-,\mathbf{p}_+}
   f_{e^-}(\mathbf{p}_-,t)f_{e^+}(\mathbf{p}_+,t)\,,\label{fgamma1}
\end{equation}

\begin{equation}
  \left(\frac{\partial f_{\gamma_j}(\mathbf{k}_i,t)}
  {\partial t}\right)_{\gamma_1\gamma_2\rightarrow e^-e^+}
 =-\int d\mathbf{k}_j d\mathbf{p}_-d\mathbf{p}_+
   w_{\mathbf{p}_-,\mathbf{p}_+;\mathbf{k}_1,\mathbf{k}_2}
   f_{\gamma_1}(\mathbf{k}_1,t)f_\gamma(\mathbf{k}_2,t)\,,
\end{equation}
for $i=1,~j=2$, and for $j=1,~ i=2$.

\begin{equation}
  \left(\frac{\partial f_{e^{\pm}}(\mathbf{p}_{\pm},t)}{\partial
  t}\right)_{e^-e^+\rightarrow\gamma_1\gamma_2}
 =-\int d\mathbf{p}_{\mp}d\mathbf{k}_1d\mathbf{k}_2
   w_{\mathbf{k}_1,\mathbf{k}_2;\mathbf{p}_-,\mathbf{p}_+}
   f_{e^-}(\mathbf{p}_-,t)f_{e^+}(\mathbf{p}_+,t)\,,
\end{equation}

\begin{equation}
  \left(\frac{\partial f_{e^{\pm}}(\mathbf{p}_{\pm},t)}
  {\partial t}\right)_{\gamma_1\gamma_2\rightarrow e^-e^+}
 = \int d\mathbf{p}_{\mp} d\mathbf{k}_1d\mathbf{k}_2
   w_{\mathbf{p}_-,\mathbf{p}_+;\mathbf{k}_1,\mathbf{k}_2}
   \frac{f_\gamma(\mathbf{k}_1,t)f_\gamma(\mathbf{k}_2,t)}{2}\,,
   \label{fe+}
\end{equation}
where
\begin{equation}
  w_{\mathbf{p}_-,\mathbf{p}_+;\mathbf{k}_1,\mathbf{k}_2}
  =\frac{ c\delta(\epsilon_-+\epsilon_+-\epsilon_1-\epsilon_2)
   }
   {(2\pi\hbar)^2}
   \delta(\mathbf{p}_-+\mathbf{p}_+-\mathbf{k}_1-\mathbf{k}_2)
   \frac{|M_{fi}|^2}
   {16\epsilon_-\epsilon_+\epsilon_1\epsilon_2},
\end{equation}
$f_{e^+}=f_{e^-}=f_e/2$ and $f_{\gamma_{1,2}}=f_\gamma/2$. Here, the
matrix element $|M_{fi}|^2$ is given by equation
(\ref{M_fi_gamma}) with the new invariants
$s =(\mathfrak{p}_--\mathfrak{k}_1)^2$ and
$u=(\mathfrak{p}_--\mathfrak{k}_2)^2$ (Berestetskii et al. 1982).

The energies of photons created via annihilation of a $e^\pm$ pair are
\begin{equation}
\epsilon_1(\mathbf{b}_1)
 =\frac{\displaystyle m_e^2c^4
        +\epsilon_-\epsilon_+(1-\beta_-\beta_+
        \mathbf{b}_-\mathbf{\cdot}\mathbf{b}_+)}
  {\displaystyle \epsilon_-(1-\beta_-\mathbf{b}_-\mathbf{\cdot}\mathbf{b}_1)
   +\epsilon_+(1-\beta_+\mathbf{b}_+\mathbf{\cdot}\mathbf{b}_1)
  }\,,\,\,\,\,\,\,\,\,\,\,\,\,\,
  \epsilon_2(\mathbf{b}_1)=\epsilon_-+\epsilon_+-\epsilon_1\,,
  \end{equation}
while the energies of pair particles created by two photons are
\begin{equation}
\epsilon_-(\mathbf{b}_-)
 =\frac{\displaystyle B\mp\sqrt{B^2-AC}}{\displaystyle A}\,,
\,\,\,\,
  \,\,\,\,\,\,\,\,\,\,\epsilon_+(\mathbf{b}_-)
 =\epsilon_1+\epsilon_2-\epsilon_-\,, \label{A27}
 \end{equation}
 where
 $ A=(\epsilon_1+\epsilon_2)^2
  -[(\epsilon_1\mathbf{b}_1+\epsilon_2\mathbf{b}_2)\mathbf{\cdot}
  \mathbf{b}_-]^2
 $,
 $B=(\epsilon_1+\epsilon_2)\epsilon_1\epsilon_2
    (1-\mathbf{b}_1\mathbf{\cdot}\mathbf{b}_2)$,
 $C= m_e^2c^4[( \epsilon_1\mathbf{b}_1
               +\epsilon_2\mathbf{b}_2)\mathbf{\cdot}\mathbf{b}_-]^2
    +\epsilon_1^2\epsilon_2^2
     (1-\mathbf{b}_1\mathbf{\cdot}\mathbf{b}_2)^2$.
The sign in equation (\ref{A27}) has to be chosen so that momentum
is conserved in the reaction.

Integration of equations (\ref{fgamma1})-(\ref{fe+}) yields
\begin{equation}
  \eta_{\gamma,\omega,k}^{e^-e^+\rightarrow\gamma_1\gamma_2}
 =\frac{1}{\Delta\epsilon_{\gamma,\omega}\Delta\mu_k}
  \left(
  \int_{\epsilon_1\in\Delta\epsilon_{\gamma,\omega}
        \atop
        \mu_1\in\Delta\mu_{k}}
  dn_{e^-}dn_{e^+}do_1
  \frac{\epsilon_1^2
         |\tilde M_{fi}|^2}
   {16\epsilon_-\epsilon_+
    \epsilon_2}
 +\int_{\epsilon_2\in\Delta\epsilon_{\gamma,\omega}
        \atop
        \mu_2\in\Delta\mu_{k}}
  dn_{e^-}dn_{e^+}do_1
  \frac{\epsilon_1
         |\tilde M_{fi}|^2}
   {16\epsilon_-\epsilon_+}
  \right),\label{A28}
\end{equation}
\begin{equation}
  (\chi E)_{e,\omega,k}^{e^-e^+\rightarrow\gamma_1\gamma_2}
 =\frac{1}{\Delta\epsilon_{e,\omega}\Delta\mu_k}
  \left(\int_{\epsilon_-\in\Delta\epsilon_{e,\omega}
        \atop
        \mu_-\in\Delta\mu_{k}}
  dn_{e^-}dn_{e^+}do_1
   \frac{\epsilon_1
         |\tilde M_{fi}|^2}
   {16\epsilon_+
    \epsilon_2}
 +\int_{\epsilon_+\in\Delta\epsilon_{e,\omega}
        \atop
        \mu_+\in\Delta\mu_{k}}
  dn_{e^-}dn_{e^+}do_1
  \frac{\epsilon_1
         |\tilde M_{fi}|^2}
   {16\epsilon_-
    \epsilon_2}
  \right),\label{A29}
\end{equation}

\begin{equation}
  (\chi E)_{\gamma,\omega,k}^{\gamma_1\gamma_2\rightarrow e^-e^+}
 =\frac{2}{\Delta\epsilon_{\gamma,\omega}\Delta\mu_k}
  \left(
  \int_{\epsilon_1\in\Delta\epsilon_{\gamma,\omega}
        \atop
        \mu_1\in\Delta\mu_{k}}
  dn_{\gamma_1} dn_{\gamma_2} do_-
  \frac{\epsilon_-\beta_-
         |\tilde M_{fi}|^2}
   {16\epsilon_2
    \epsilon_+}
 +\int_{\epsilon_2\in\Delta\epsilon_{\gamma,\omega}
        \atop
        \mu_2\in\Delta\mu_{k}}
  dn_{\gamma_1} dn_{\gamma_2} do_-
  \frac{\epsilon_-\beta_-
         |\tilde M_{fi}|^2}
   {16\epsilon_1
    \epsilon_+}
  \right),\label{A30}
\end{equation}
\begin{equation}
  \eta_{e,\omega,k}^{\gamma_1\gamma_2\rightarrow e^-e^+}
 =\frac{2}{\Delta\epsilon_{e,\omega}\Delta\mu_k}
  \left(
  \int_{\epsilon_-\in\Delta\epsilon_{e,\omega}
        \atop
        \mu_-\in\Delta\mu_{k}}
  dn_{\gamma_1} dn_{\gamma_2} do_-
  \frac{\epsilon_-^2\beta_-
  |\tilde M_{fi}|^2}
    {16\epsilon_1\epsilon_2
    \epsilon_+}
 +\int_{\epsilon_+\in\Delta\epsilon_{e,\omega}
        \atop
        \mu_+\in\Delta\mu_{k}}
  dn_{\gamma_1} dn_{\gamma_2} do_-
  \frac{\epsilon_-\beta_-
  |\tilde M_{fi}|^2}
    {16\epsilon_1\epsilon_2}
  \right),
\label{A31}
\end{equation}
where $dn_{e^\mp}=d\epsilon_\mp do_\mp \epsilon_\mp^2\beta_\mp
f_{e^\mp}$, and $dn_{\gamma_{1,2}}=d\epsilon_{1,2} do_{1,2}
  \epsilon_{1,2}^2 f_{\gamma_{1,2}}$.

\subsection{M{\o}ller scattering of electrons and positrons}
The time evolution of the distribution functions of electrons (or
positrons) due to M{\o}ller scattering,
$e^{\pm}e^{\pm}\rightarrow e^{\pm}e^{\pm}$, are
described by
\begin{equation}
  \left(\frac{\partial f_{e_i}(\mathbf{p}_i,t)}
  {\partial t}\right)_{e_1e_2\rightarrow e_1'e_2'}
 =\int d\mathbf{p}_j d\mathbf{p}_1'd\mathbf{p}_2'
  w_{\mathbf{p}_1',\mathbf{p}_2';\mathbf{p}_1,\mathbf{p}_2}
  [ f_{e_1}(\mathbf{p}_1',t)f_{e_2}(\mathbf{p}_2',t)
   -f_{e_1}(\mathbf{p}_1,t)f_{e_2}(\mathbf{p}_2,t)]\,,\label{A32}
\end{equation}
with $i=1,~j=2$, and with $j=1,~i=2$, and where
\begin{equation}
  w_{\mathbf{p}_1',\mathbf{p}_2';\mathbf{p}_1,\mathbf{p}_2}
  =\frac{c\delta(\epsilon_1+\epsilon_2-\epsilon_1'-\epsilon_2')
   }
   {(2\pi\hbar)^2}
   \delta(\mathbf{p}_1+\mathbf{p}_2-\mathbf{p}_1'-\mathbf{p}_2')
   \frac{|M_{fi}|^2}
   {16\epsilon_1\epsilon_2\epsilon_1'\epsilon_2'},
\end{equation}
\begin{eqnarray}
  |M_{fi}|^2
      =2^6(\pi\hbar)^2r_e^2m_e^2c^7
       \Biggl\{
          \frac{1}{g^2}
          \left[ \frac{s^2+u^2}{2}
                +4m_e^2c^2(g-m_e^2c^2)
          \right]
         +\frac{1}{u^2}\left[ \frac{s^2+g^2}{2}
                             +4m_e^2c^2(u-m_e^2c^2)
                       \right]
          \nonumber \\
         +\frac{4}{g u}
          \left(\frac{s}{2}-m_e^2c^2\right)
          \left(\frac{s}{2}-3m_e^2c^2\right)
       \Biggr\}, \label{M_fi_e}
\end{eqnarray}
$f_{e_{1}}({\bf p}_1,t)=f_e({\bf p}_1,t)/2$,
$f_{e_{2}}({\bf p}_2,t)=f_e({\bf p}_2,t)/2$,
$s =(\mathfrak{p}_1+\mathfrak{p}_2)^2
 =2(m_e^2c^2+\mathfrak{p}_1\mathbf{\cdot}\mathfrak{p}_2)$,
$g
 =(\mathfrak{p}_1-\mathfrak{p}_1')^2
 =2(m_e^2c^2-\mathfrak{p}_1\mathbf{\cdot}\mathfrak{p}_1')$, and
$u
 =(\mathfrak{p}_1-\mathfrak{p}_2')^2
=2(m_e^2c^2-\mathfrak{p}_1\mathbf{\cdot}
 \mathfrak{p}_2')$ (Berestetskii et al. 1982).

The energies of final-state particles are
\begin{equation}
\epsilon_1'(\mathbf{b}_1')
 =\tilde B\mp\frac{\displaystyle\sqrt{\tilde B^2-\tilde A \tilde C}}
 {\displaystyle \tilde A}\,,
 \,\,\,\,\,\,\,\,\,\,
\epsilon_2'(\mathbf{b}_1')
 =\epsilon_1+\epsilon_2-\epsilon_1'\,,
 \end{equation}
 where $\tilde A=(\epsilon_1+\epsilon_2)^2
  -( \epsilon_1\beta_1\mathbf{b}_1\mathbf{\cdot}\mathbf{b}_1'
    +\epsilon_2\beta_2\mathbf{b}_2\mathbf{\cdot}\mathbf{b}_1')^2
$,
 $\tilde B=(\epsilon_1+\epsilon_2)
    [ m_e^2c^4
     +\epsilon_1\epsilon_2(1-\beta_1\beta_2\mathbf{b}_1
     \mathbf{\cdot}\mathbf{b}_2)
    ]$, and
 $\tilde C= m_e^2c^4( \epsilon_1\beta_1\mathbf{b}_1\mathbf{\cdot}\mathbf{b}_1'
              +\epsilon_2\beta_2\mathbf{b}_2\mathbf{\cdot}\mathbf{b}_1')^2
    +[ m_e^2c^4
      +\epsilon_1\epsilon_2
       (1-\beta_1\beta_2\mathbf{b}_1\mathbf{\cdot}\mathbf{b}_2)
     ]^2$.

Intergration of equations (\ref{A32}),  similar to the case of
Compton scattering, yields
\begin{equation}
  \eta_{e,\omega,k}^{e_1e_2\rightarrow e_1'e_2'}
 =\frac{1}{\Delta\epsilon_{e,\omega}\Delta\mu_k}
  \Biggl(
  \int_{\epsilon_1'\in\Delta\epsilon_{e,\omega}
        \atop
        \mu_1'\in\Delta\mu_{k}}
  dn_{e_1}dn_{e_2} do_1'
   \frac{\epsilon_1'^2\beta_1'
   |\tilde M_{fi}|^2}
   {16\epsilon_1\epsilon_2
    \epsilon_2'}
  +\int_{\epsilon_2'\in\Delta\epsilon_{e,\omega}
        \atop
        \mu_2'\in\Delta\mu_{k}}
  dn_{e_1}dn_{e_2} do_1'
   \frac{\epsilon_1'\beta_1'
         |\tilde M_{fi}|^2}
   {16\epsilon_1\epsilon_2}
  \Biggr),
\end{equation}
\begin{equation}
  (\chi E)_{e,\omega,k}^{e_1e_2\rightarrow e_1'e_2'}
 =\frac{1}{\Delta\epsilon_{e,\omega}\Delta\mu_k}
  \Biggl(
  \int_{\epsilon_1\in\Delta\epsilon_{e,\omega}
        \atop
        \mu_1\in\Delta\mu_{k}}
  dn_{e_1}dn_{e_2} do_1'
   \frac{\epsilon_1'\beta_1'
         |\tilde M_{fi}|^2}
   {16\epsilon_2
    \epsilon_2'}
 +\int_{\epsilon_2\in\Delta\epsilon_{e,\omega}
        \atop
        \mu_2\in\Delta\mu_{k}}
  dn_{e_1}dn_{e_2} do_1'
   \frac{\epsilon_1'\beta_1'
         |\tilde M_{fi}|^2}
   {16\epsilon_1
    \epsilon_2'}
  \Biggr),
\end{equation}
where
$dn_{e_{1,2}}=d\epsilon_{1,2}do_{1,2}
  \epsilon_{1,2}^2\beta_{1,2} f_{e_{1,2}}$.

\subsection{Bhaba scattering of electrons on positrons}
The time evolution of the distribution functions of electrons and
positrons due to Bhaba scattering, $e^-e^+\rightarrow
e^{-\prime}e^{+\prime}$, is described by

\begin{equation}
  \left(\frac{\partial f_{e^{\pm}}(\mathbf{p}_{\pm},t)}
  {\partial t}\right)_{e^-e^+\rightarrow
e^{-\prime}e^{+\prime}}
 =\int d\mathbf{p}_{\mp}d\mathbf{p}_-'d\mathbf{p}_+'
  w_{\mathbf{p}_-',\mathbf{p}_+';\mathbf{p}_-,\mathbf{p}_+}
  [ f_{e^-}(\mathbf{p}_-',t)f_{e^+}(\mathbf{p}_+',t)
   -f_{e^-}(\mathbf{p}_-,t)f_{e^+}(\mathbf{p}_+,t)],\label{A38}
\end{equation}

where
\begin{equation}
  w_{\mathbf{p}_-',\mathbf{p}_+';\mathbf{p}_-,\mathbf{p}_+}
  =\frac{c\delta(\epsilon_-+\epsilon_+-\epsilon_-'-\epsilon_+')
   }
   {(2\pi\hbar)^2}
   \delta(\mathbf{p}_-+\mathbf{p}_+-\mathbf{p}_-'-\mathbf{p}_+')
   \frac{|M_{fi}|^2}
   {16\epsilon_-\epsilon_+\epsilon_-'\epsilon_+'},
\end{equation}
$|M_{fi}|$ is given by equation (\ref{M_fi_e}), but the invariants
are $s =(\mathfrak{p}_--\mathfrak{p}_+')^2$,
$g=(\mathfrak{p}_+-\mathfrak{p}_+')^2$ and
$u=(\mathfrak{p}_-+\mathfrak{p}_+)^2$. The final energies
$\epsilon_-'$, $\epsilon_+'$ are functions of the outgoing
particle directions in a way similar to that in Section A.3 (see
Berestetskii et al. 1982).

Integration of equations (\ref{A38}) yields
\begin{equation}
  \eta_{e,\omega,k}^{e^-e^+\rightarrow e^{-\prime}e^{+\prime}}
 =\frac{1}{\Delta\epsilon_{e,\omega}\Delta\mu_k}
  \Biggl(\int_{\epsilon_-'\in\Delta\epsilon_{e,\omega}
        \atop
        \mu_-'\in\Delta\mu_{k}}
  dn_{e^-}dn_{e^+}do_-'
   \frac{\epsilon_-'^2\beta_-'
         |\tilde M_{fi}|^2}
   {16\epsilon_-\epsilon_+
    \epsilon_+'}
 +\int_{\epsilon_+'\in\Delta\epsilon_{e,\omega}
        \atop
        \mu_+'\in\Delta\mu_{k}}
  dn_{e^-}dn_{e^+}do_-'
   \frac{\epsilon_-'\beta_-'
         \tilde |M_{fi}|^2}
   {16\epsilon_-\epsilon_+}
  \Biggr),
\end{equation}
\begin{equation}
  (\chi E)_{e,\omega,k}^{e^-e^+\rightarrow e^{-\prime}e^{+\prime}}
 =\frac{1}{\Delta\epsilon_{e,\omega}\Delta\mu_k}
  \Biggl(
  \int_{\epsilon_-\in\Delta\epsilon_{e,\omega}
        \atop
        \mu_-\in\Delta\mu_{k}}
  dn_{e^-}dn_{e^+}do_-'
   \frac{\epsilon_-'\beta_-'
         |\tilde M_{fi}|^2}
   {16\epsilon_+
    \epsilon_+'}
 +\int_{\epsilon_+\in\Delta\epsilon_{e,\omega}
        \atop
        \mu_+\in\Delta\mu_{k}}
  dn_{e^-}dn_{e^+}do_-'
   \frac{\epsilon_-'\beta_-'
         |\tilde M_{fi}|^2}
   {16\epsilon_-
    \epsilon_+'}
  \Biggr),
\end{equation}
where $dn_{e^\mp}=d\epsilon_\mp do_\mp \epsilon_\mp^2\beta_\mp
f_{e^\mp}$.

\subsection{Radiative three-body processes}
Compton scattering and scattering of $e^\pm$ pairs can equalize
the temperatures of photons, electrons, and positrons. In the
absence of three-body processes, the total particle number is
conserved, and the equilibrium distribution function of the
photons need not take the Planck form. The equilibrium of the
reaction $e^+e^-\rightleftarrows \gamma_1\gamma_2$ leads to
equality
of the chemical potentials of pairs and photons,
$\varphi_e=\varphi_\gamma$.

When only two-body processes are included $\varphi_\gamma\not
=0$, in general. To get more realistic spectra we should include
reactions that don't conserve particle number. The rates of such
three-body reactions are at least $\alpha^{-1}\sim 10^2$ times
smaller than the rates of the two-particle reactions. The
three-body processes we include in our study are listed in
Table~1. We adopt the following emission coefficients for these
reactions (cf. Haug, 1985; Svensson, 1984).
\begin{enumerate}
\item Bremsstrahlung, $e e\rightarrow e e \gamma$
\begin{equation}
  \eta_{\gamma}^{e^\mp e^\mp\rightarrow e^\mp e^\mp \gamma}
  =(n_{e^+}^2+n_{e^-}^2)\frac{8c r_e^2}{411}
   \ln\left[4\xi(11.2+10.4\theta^2)
                        \left(1+\frac{\theta}{x}\right)
                  \right]
   \left(\frac{3}{5}\sqrt{2}\theta+2\theta^2\right)
   \frac{\exp(-x/\theta)}{\exp(1/\theta)K_2(1/\theta)},
 \end{equation}
 \begin{equation}
  \eta_{\gamma}^{e^- e^+\rightarrow e^- e^+ \gamma}
  =n_{e^+}n_{e^-}\frac{8c r_e^2}{411}
   \ln\left[4\xi(1+10.4\theta)
                        \left(1+\frac{\theta}{x}\right)
                  \right]
   2(\sqrt{2}+2\theta+2\theta^2)
   \frac{\exp(-x/\theta)}{\exp(1/\theta)K_2(1/\theta)},
   \end{equation}
where $x=\epsilon/(m_e c^2)$, $\xi=e^{-0.5772}$,
$\theta=k_\mathrm{B}T_e/(m_e c^2)$, and $K_2(1/\theta)$ is the
modified Bessel function of the second kind of order 2. The
electron temperature, $T_e$, for the energy density $E_e$ may be
determined from
\begin{equation}
\int E_e d\mu d\epsilon=\int E_e^\mathrm{eq} d\mu d\epsilon\,,
\end{equation}
where $E_e^{\rm eq}=2\pi\epsilon^3f_e^{\rm eq}/c^3$,
\begin{equation}
  f_e^\mathrm{eq}(\epsilon)=\frac{2}{(2\pi\hbar)^3}\exp
  \left(-\frac{\epsilon-\varphi_e}{kT}\right). \label{eqfunctions}
\end{equation}
is the equilibrium distribution function for electrons and positrons
(degeneracy of particles is neglected), and $\varphi_e$ is the
chemical potential.

\item Double Compton scattering, $\gamma e\rightarrow \gamma e
\gamma$
\begin{equation}
  \eta_\mathrm{\gamma}^{\gamma e\rightarrow \gamma e\gamma}
  =(n_{e^+}+n_{e^-})n_\gamma\frac{64c r_e^2}{411}
  \frac{\theta^2\exp(-x/\theta)}
   {1+13.91\theta+11.05\theta^2+19.92\theta^3}.
\end{equation}
\item Three photon annihilation, $e^-e^+\rightarrow \gamma\gamma\gamma$
\begin{equation}
  \eta_\gamma^{e^-e^+\rightarrow \gamma \gamma \gamma}
  =n_{e^+}n_{e^-}
   \frac{2c r_e^2}{137}
   \theta\exp(-x/\theta).
\end{equation}
\end{enumerate}

We use the absorption coefficient for three-body processes
written as
\begin{equation}
\chi_\gamma^{\rm 3p}
=\eta_\gamma^\mathrm{3p}/E_\gamma^\mathrm{eq}\,,
\end{equation}
where $\eta_\gamma^\mathrm{3p}$ is the sum of
the emission coefficients of photons in the three particle processes,
$E_\gamma^{\rm eq}=2\pi\epsilon^3f_\gamma^{\rm eq}/c^3$, and
\begin{equation}
f_\gamma (\epsilon)^\mathrm{eq}=\frac{\displaystyle 2}{\displaystyle
(2\pi\hbar)^3[\exp(\epsilon/kT) -1]}
\end{equation}
is the equilibrium distribution function for photons.

From equation (\ref{Boltzmann}), the law of energy conservation in
the three-body processes is
\begin{equation}
\int{\sum_i(\eta^{\rm 3p}_i -\chi_i^{\rm 3p}E_i)d\mu d\epsilon}=0\,.
\end{equation}
For exact conservation of energy in these processes we introduce the
following coefficients of emission and absorption for electrons:
\begin{eqnarray}
  \chi_e^{\rm 3p}=\frac{\int(\eta_\gamma^{\rm 3p}-
  \chi_\gamma^{\rm 3p}E_\gamma)d\epsilon d\mu}
              {\int E_ed\epsilon d\mu},\mbox{ }\,\,\,\,\,\,\,\,
  \eta_e^{\rm 3p}=0,\,\,\,\,\,\,\,\,\,\,\,\,\,\mbox{ if}
  \int(\eta_\gamma^{\rm 3p}-\chi_\gamma^{\rm 3p}E_\gamma)
  d\epsilon d\mu>0\,,
\end{eqnarray}
or
\begin{eqnarray}
  \frac{\eta_e^{\rm 3p}}{E_e}=-\frac{\int(\eta_\gamma^{\rm 3p}-
  \chi_\gamma^{\rm 3p}E_\gamma)d\epsilon d\mu}
              {\int E_ed\epsilon d\mu},\mbox{ }\,\,\,\,\,\,\,\,
  \chi_e^{\rm 3p}=0,\,\,\,\,\,\,\,\,\,\,\,\,\,\mbox{ if}
  \int(\eta_\gamma^{\rm 3p}-\chi_\gamma^{\rm 3p}E_\gamma)d\epsilon
  d\mu<0\,.
\end{eqnarray}

\end{appendix}

\onecolumn

\begin{table*}
\caption{Physical Processes Included in Simulations}
\begin{center}
\begin{tabular}{ll}
  \hline \hline
  Basic Two-Body & Radiative \\
  Interaction & Variant \\
  \\ \hline
  M{\o}ller and Bhaba &  \\
  scattering & Bremsstrahlung \\
  $ee\rightarrow ee$ & $ee\leftrightarrow ee\gamma$ \\ \hline
  Compton scattering \,\,\,\,\,\,\,\,\,\,& Double Compton scattering \\
  $\gamma e\rightarrow \gamma e$ & $\gamma e\leftrightarrow \gamma
e\gamma$
  \\ \hline
  Pair annihilation & Three photon annihilation \\
  $e^+e^-\rightarrow \gamma\gamma$ & $e^+e^-\leftrightarrow
\gamma\gamma\gamma$
  \\ \hline
  Photon-photon &  \\
  pair production &   \\
  $\gamma\gamma\rightarrow e^+e^-$ &   \\ \hline
\end{tabular}
\end{center}
\end{table*}
\begin{figure}
\plotone{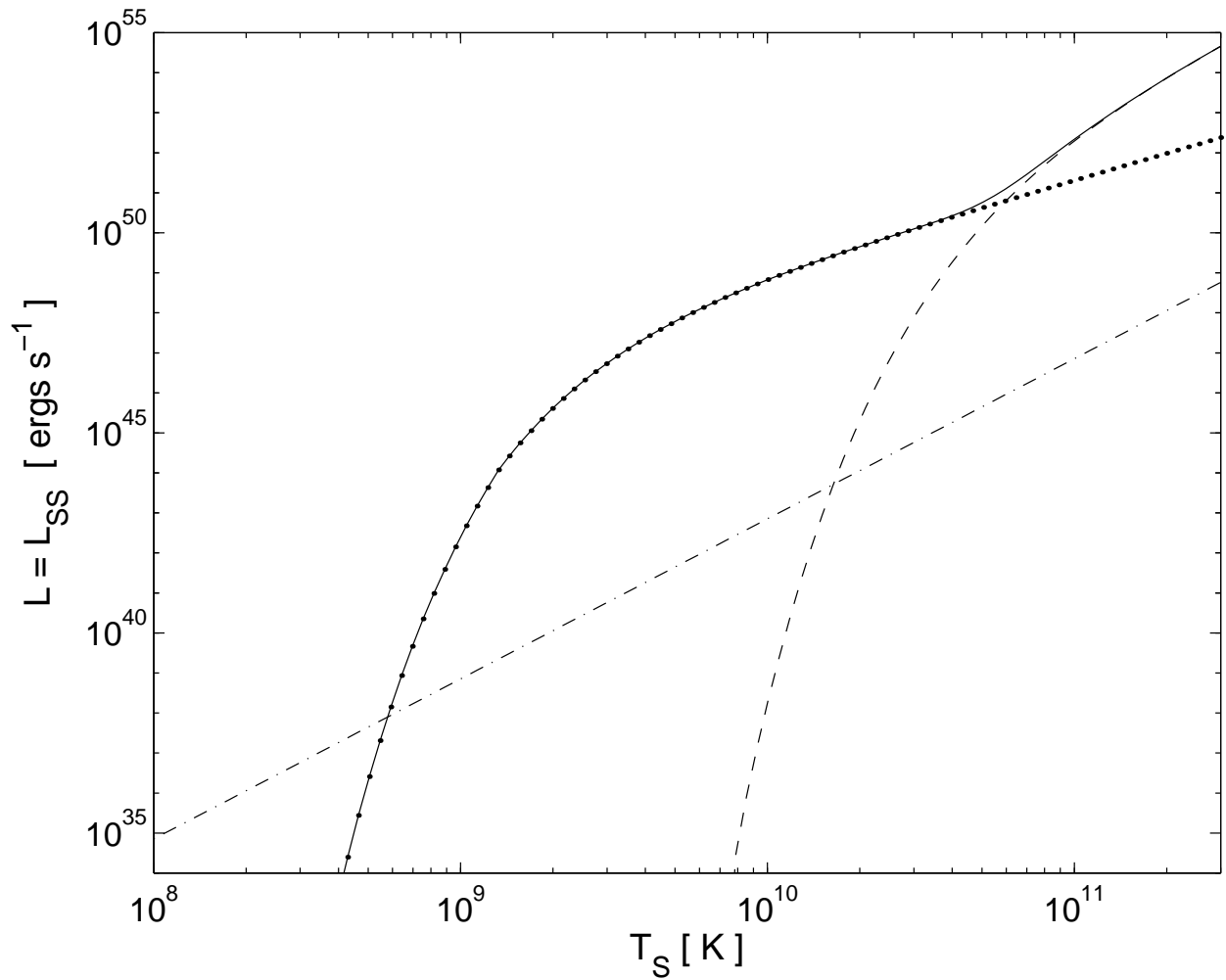} \caption{The luminosities of a hot, bare, strange
star in $e^+e^-$ pairs (dotted line), in thermal equilibrium
photons (dashed line), and the total (solid line) as functions of
the surface temperature $T_{_{\rm S}}$. The upper limit on the
luminosity in non-equilibrium photons, $L_{\rm neq}\lesssim
10^{-6}L_{\rm BB}$, is shown by the dot-dashed line, $L_{\rm BB}$
being the blackbody luminosity. } \label{Ltotal-Ts}
\end{figure}
\begin{figure}
\plotone{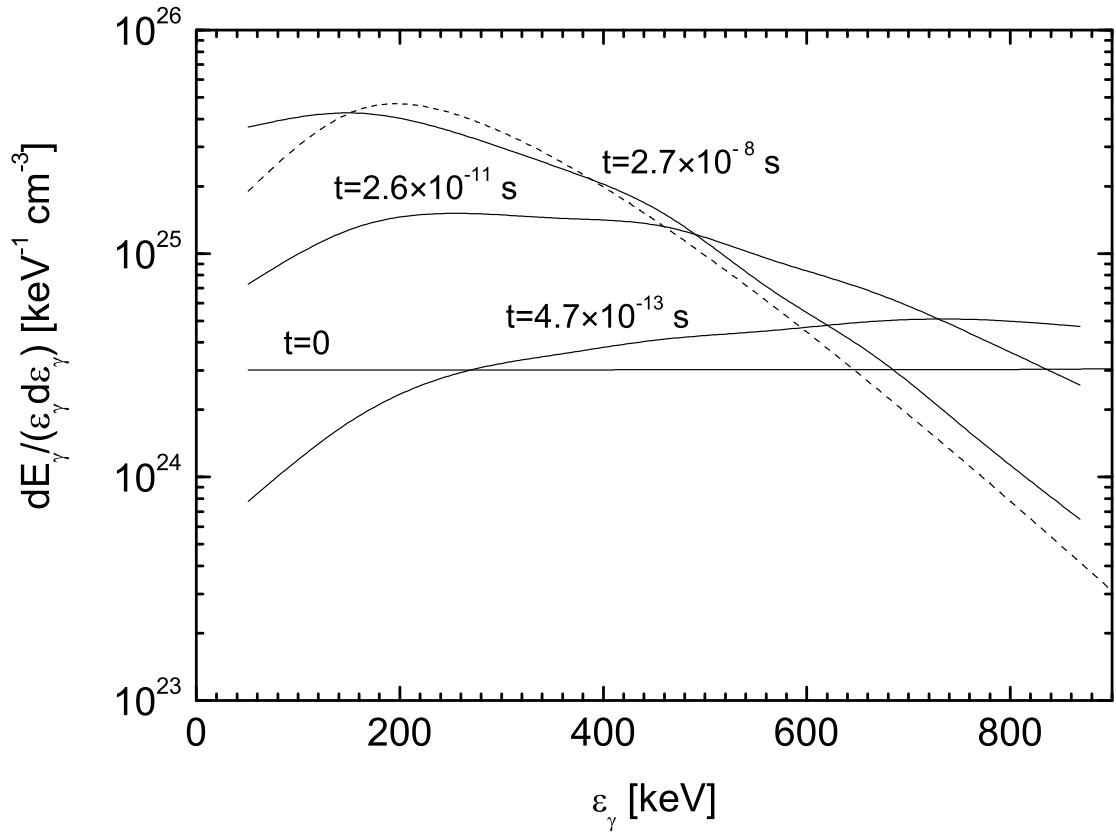} \caption{Time evolution of the photon spectrum in
the test problem of a homogeneous distribution starting away from
equilibrium. The dashed line is the black body spectrum with
the temperature of $10^9$ K. Times are marked on the curves.}
\label{Lth-time}
\end{figure}
\begin{figure}
\plotone{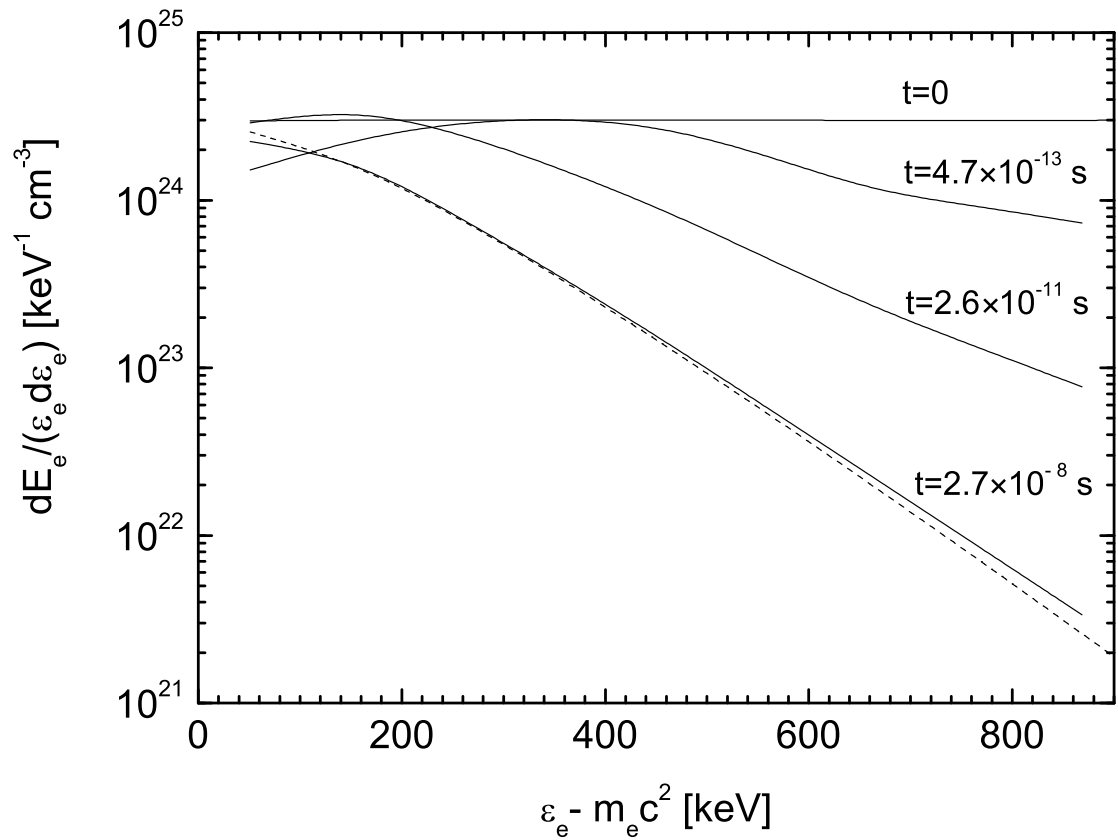} \caption{Same as Fig. 2 for pairs. The dashed line is 
the relativistic Maxwellian for the temperature of $10^9$ K.}
\label{tau.Fig}
\end{figure}
\begin{figure}
\plotone{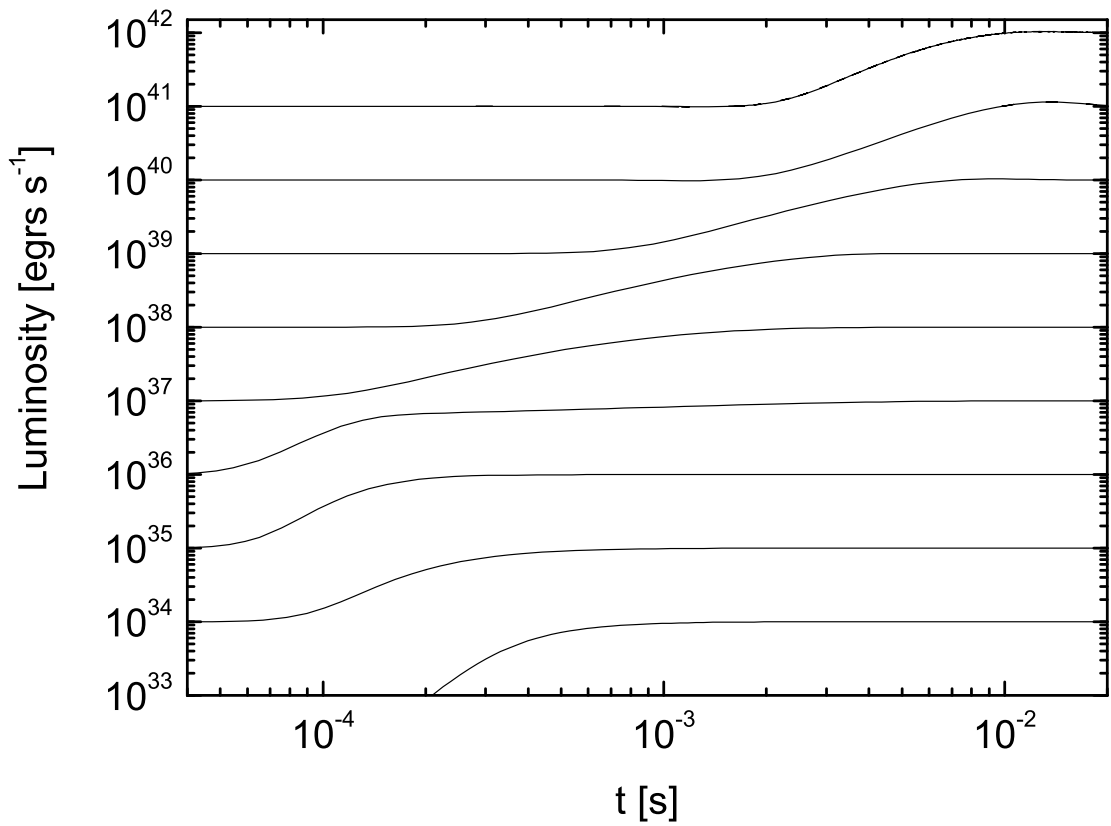} \caption{The total emerging luminosity in photons
and pairs at the external boundary ($r=r_{\rm ext}$) as a function
of time in runs with different values of $\dot{\rm E}$ (equal to
the long-time asymptotic value of $L$).} \label{Lth-time}
\end{figure}
\begin{figure}
\plotone{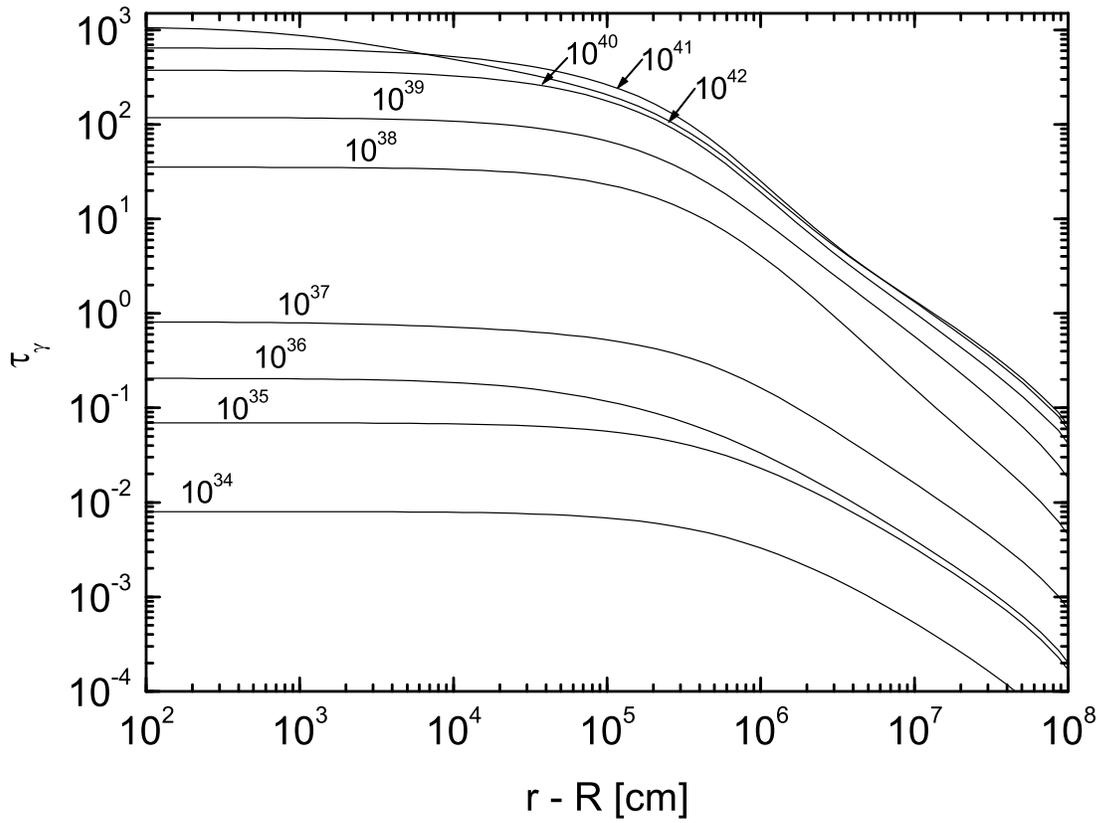} \caption{The mean optical depth for photons, from
$r$ to infinity, as a function of the distance from the stellar
surface, for different values of $\dot{\rm E}$, as marked on the
curves.} \label{tau.Fig}
\end{figure}
\begin{figure}
\plotone{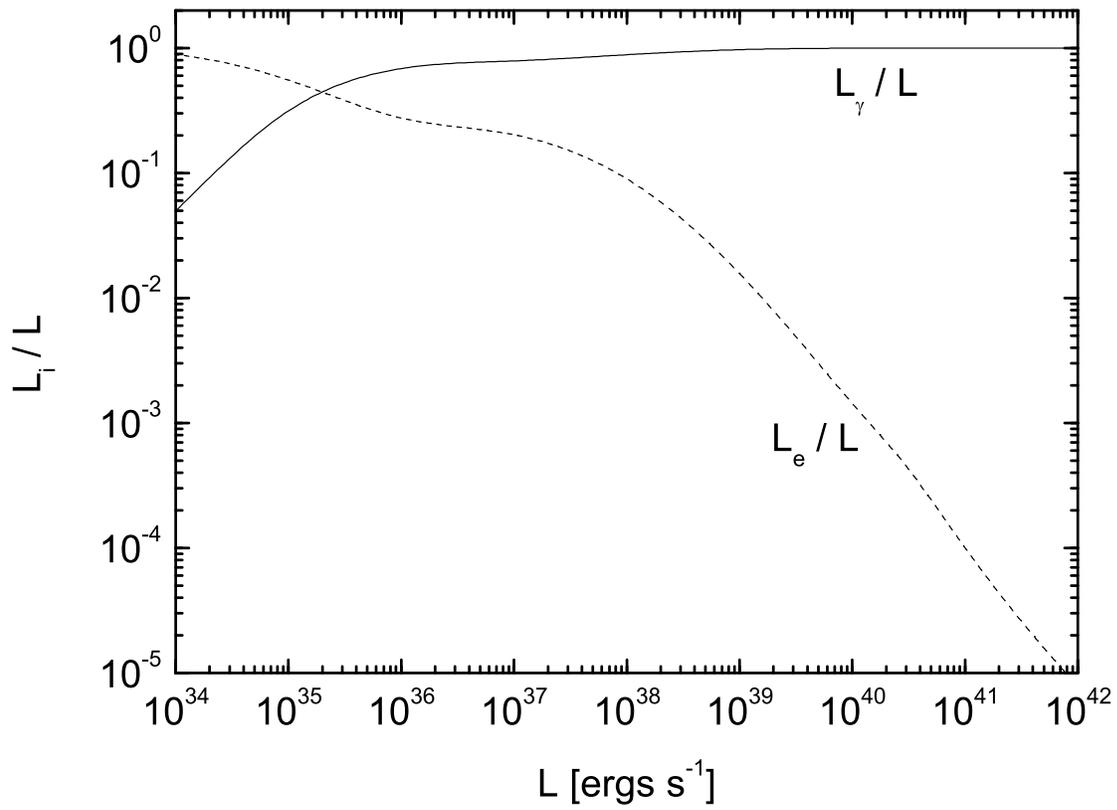} \caption{The fractional emerging luminosities in
pairs ($L_e$) and photons ($L_\gamma$) as functions of the total
luminosity, $L=L_e+L_\gamma$.} \label{Li/L.Fig}
\end{figure}
\begin{figure}
\plotone{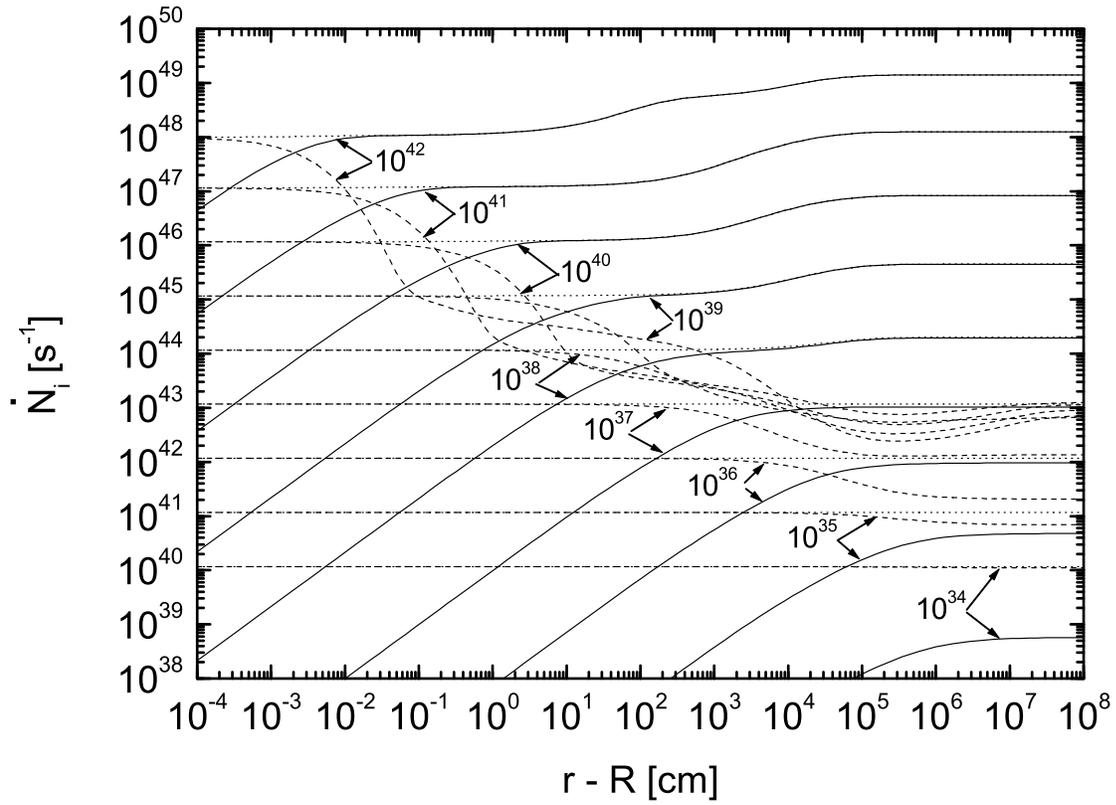} \caption{The rates of particle number outflow in
photons (solid lines), in $e^+e^-$ pairs (dashed lines), and the
total (dotted lines) through the surface at radius $r$ as a
function of the distance from the stellar surface for different
values of $\dot{\rm E}$, as marked on the curves. } \label{N.Fig}
\end{figure}

\begin{figure}
\plotone{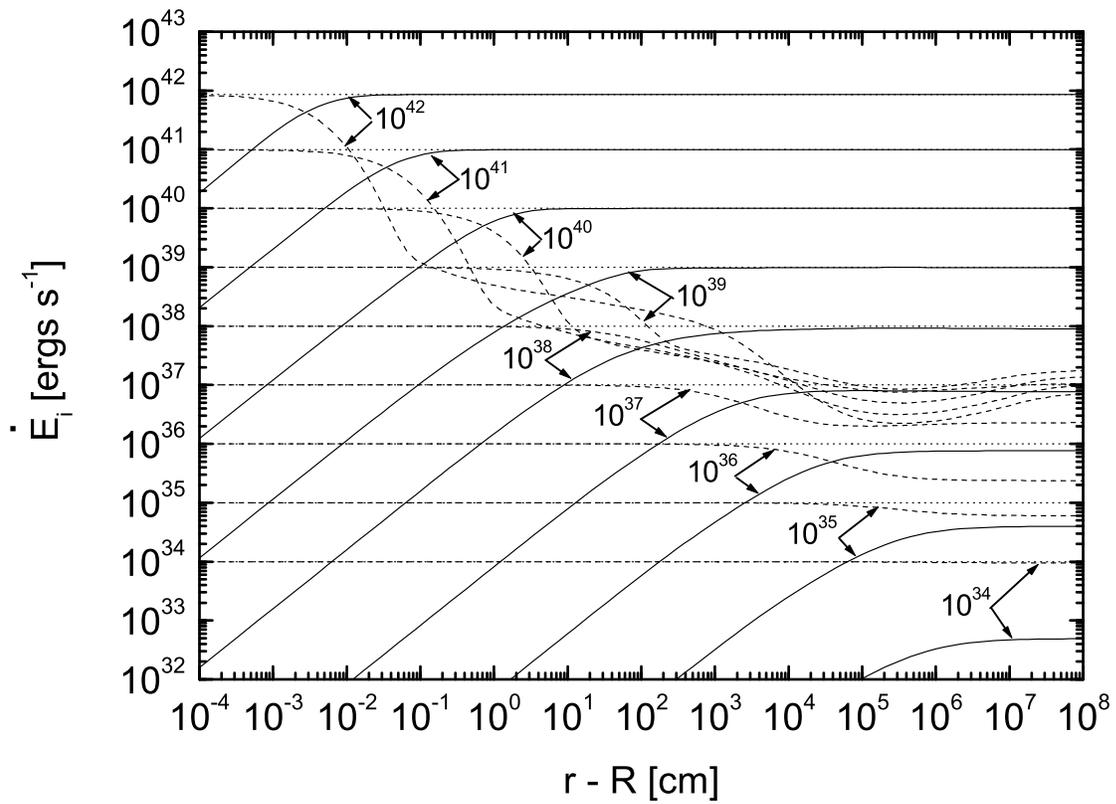} \caption{The rate of energy outflow, as in Figure
7. } \label{L.Fig}
\end{figure}
\begin{figure}
\plotone{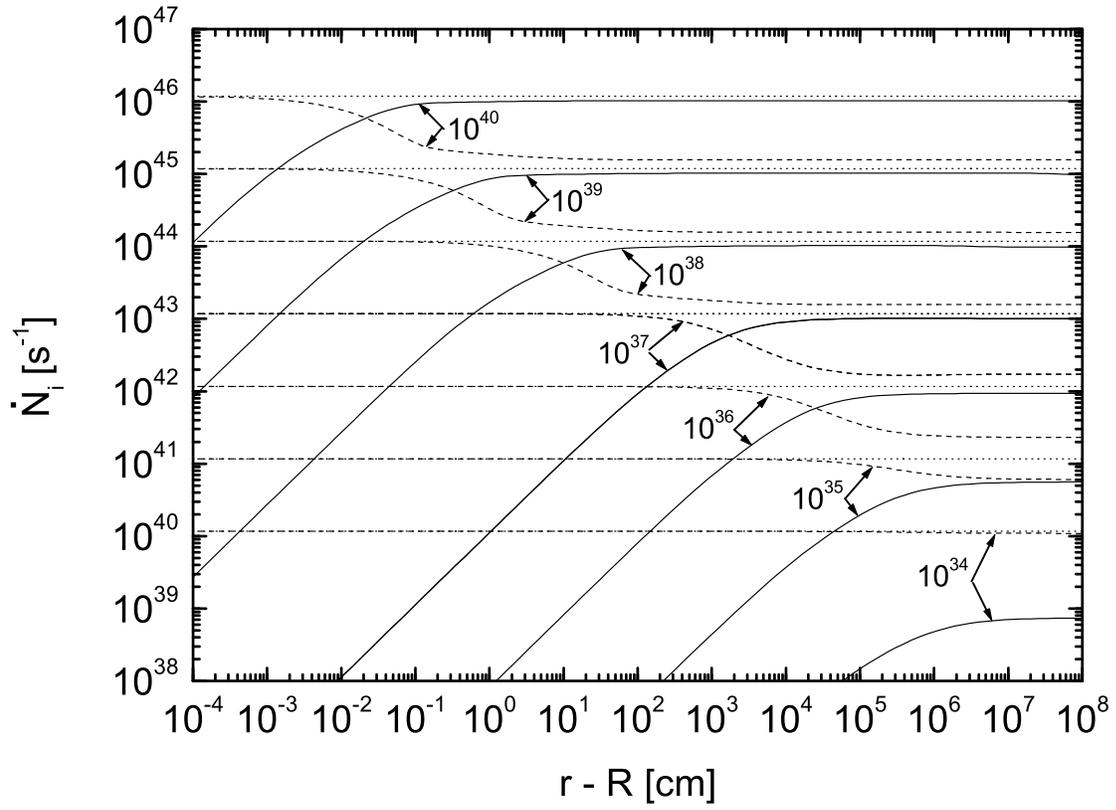} \caption{The same as Figure 7, with only
two-particle processes taken into account.} \label{L2p.Fig}
\end{figure}
\begin{figure}
\plotone{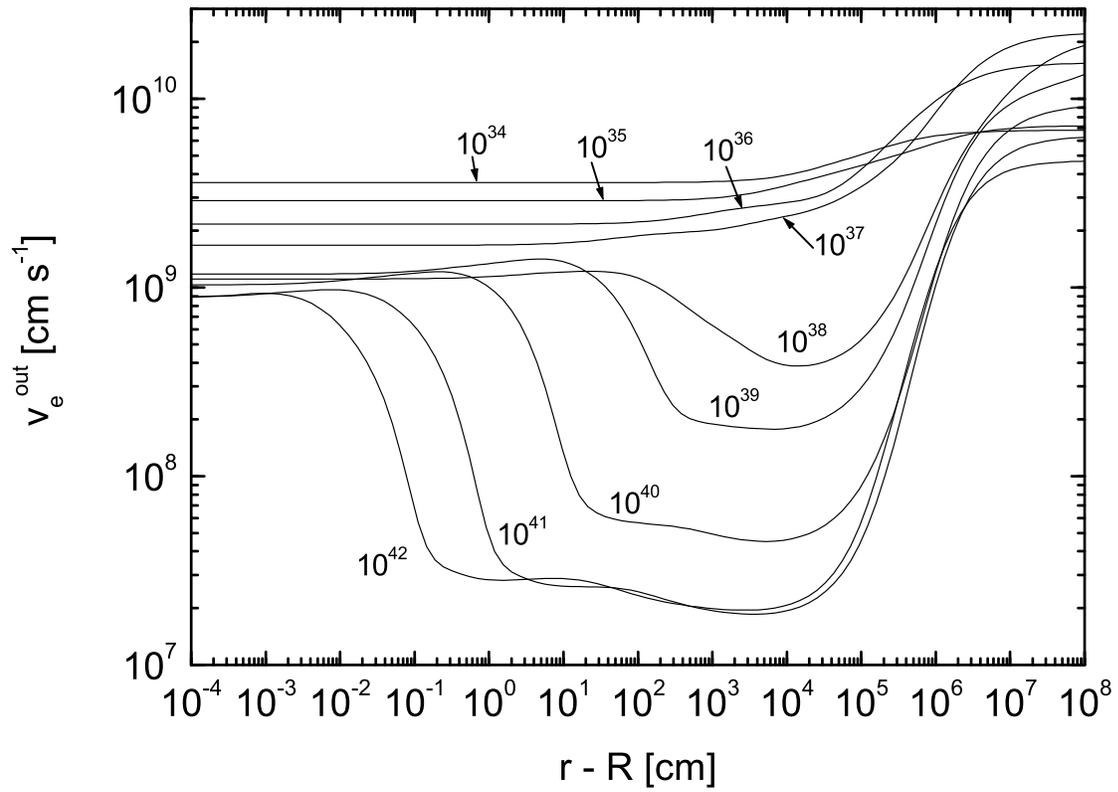} \caption{The velocity of the pair plasma outflow
as a function of the distance from the stellar surface for
different values of $\dot{\rm E}$, as marked on the curves.}
\label{v-r}
\end{figure}
\begin{figure}
\plotone{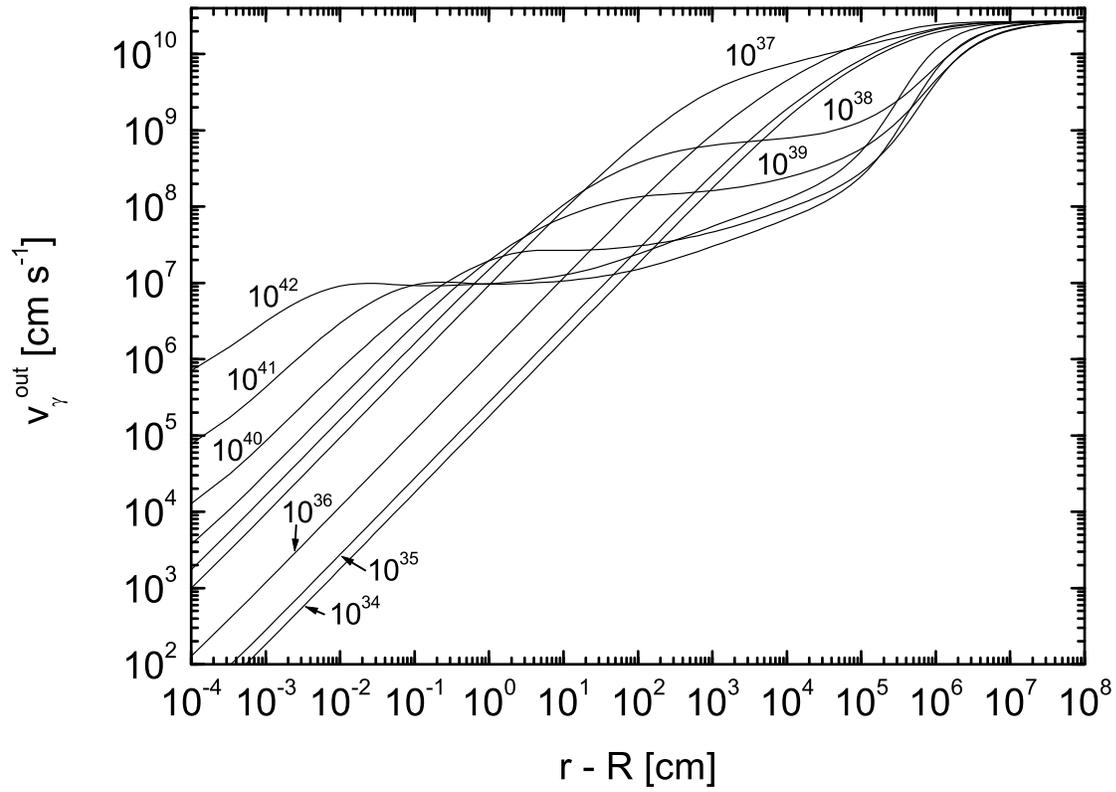} \caption{The velocity of the photon gas outflow
as a function of the distance from the stellar surface for
different values of $\dot{\rm E}$, as marked on the curves
($v_{\rm out}^\gamma/c$ is a measure of photon anisotropy).}
\label{v-gamma}
\end{figure}
\begin{figure}
\plotone{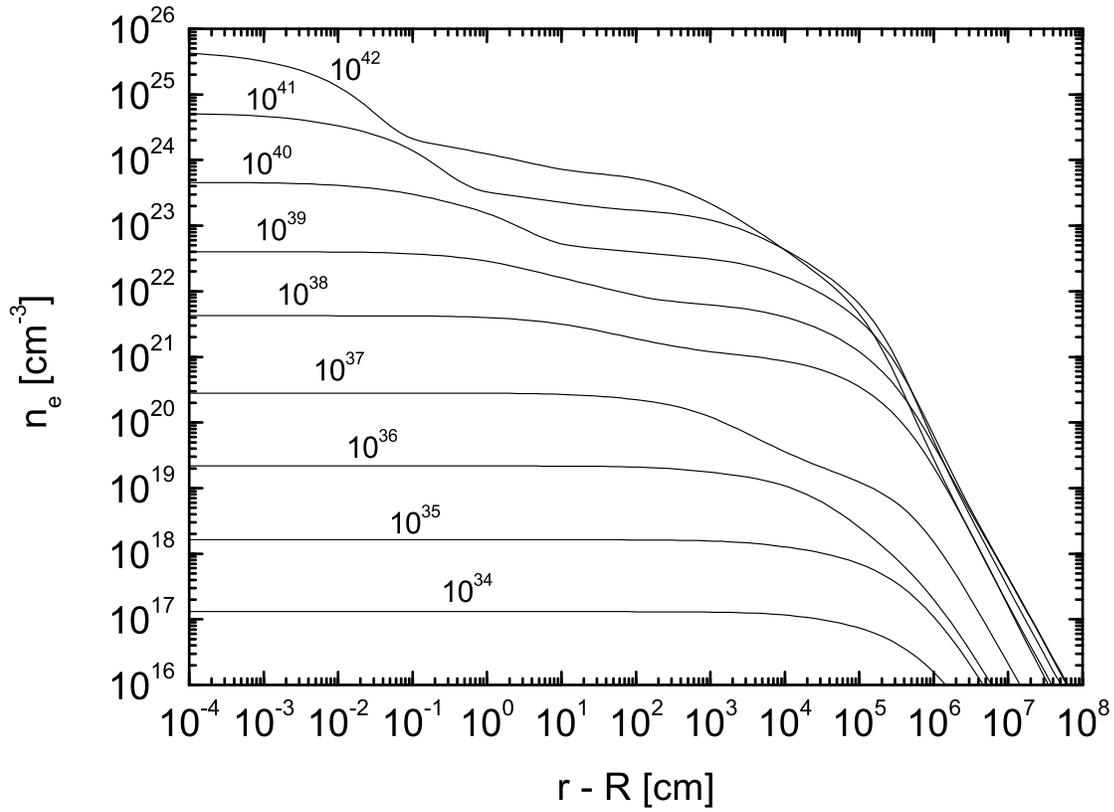} \caption{The pair number density as a function
of the distance from the stellar surface for different values of
$\dot{\rm E}$, as marked on the curves.}
\end{figure}
\begin{figure}
\plotone{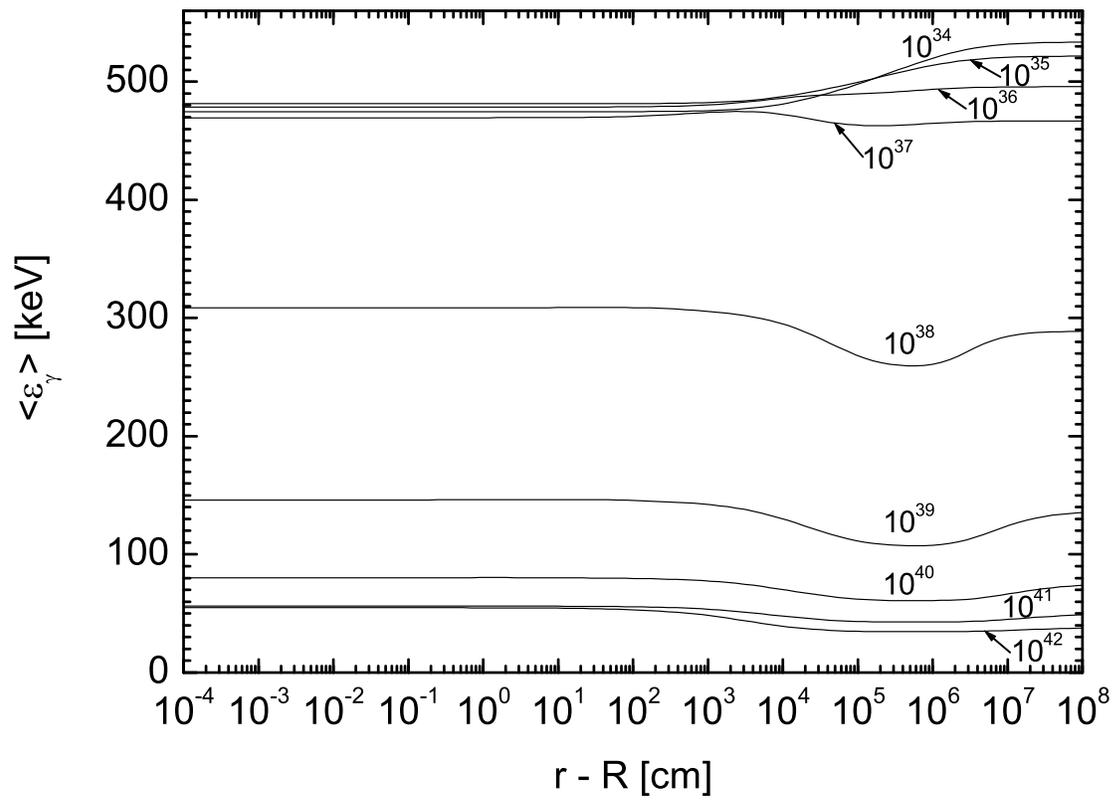} \caption{The mean photon energy as a function
of the distance from the stellar surface for different values of
$\dot{\rm E}$,
as
marked on the curves.} \label{gammaepsilon.Fig}
\end{figure}
\begin{figure}
\plotone{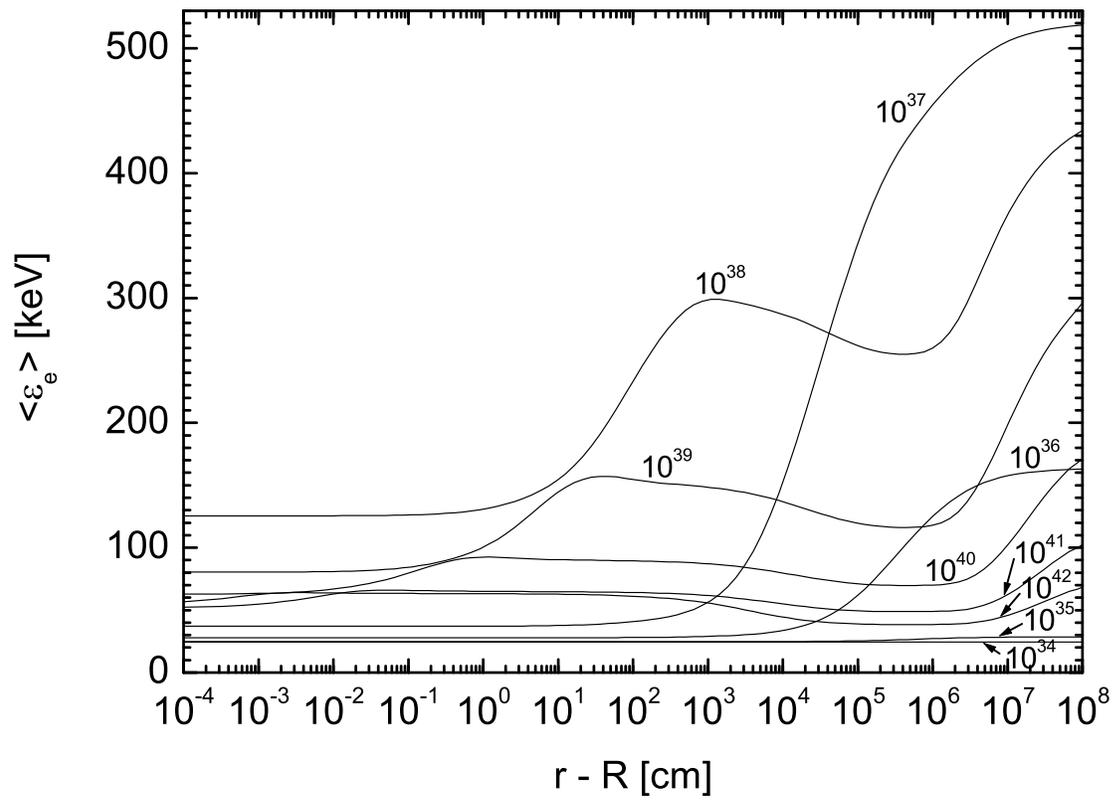} \caption{The mean kinetic energy of electrons and positrons
as a function of the distance from the stellar surface for different
values of $\dot{\rm E}$,
as
marked on the curves.}
\label{electronepsilon.Fig}
\end{figure}
\begin{figure}
\plotone{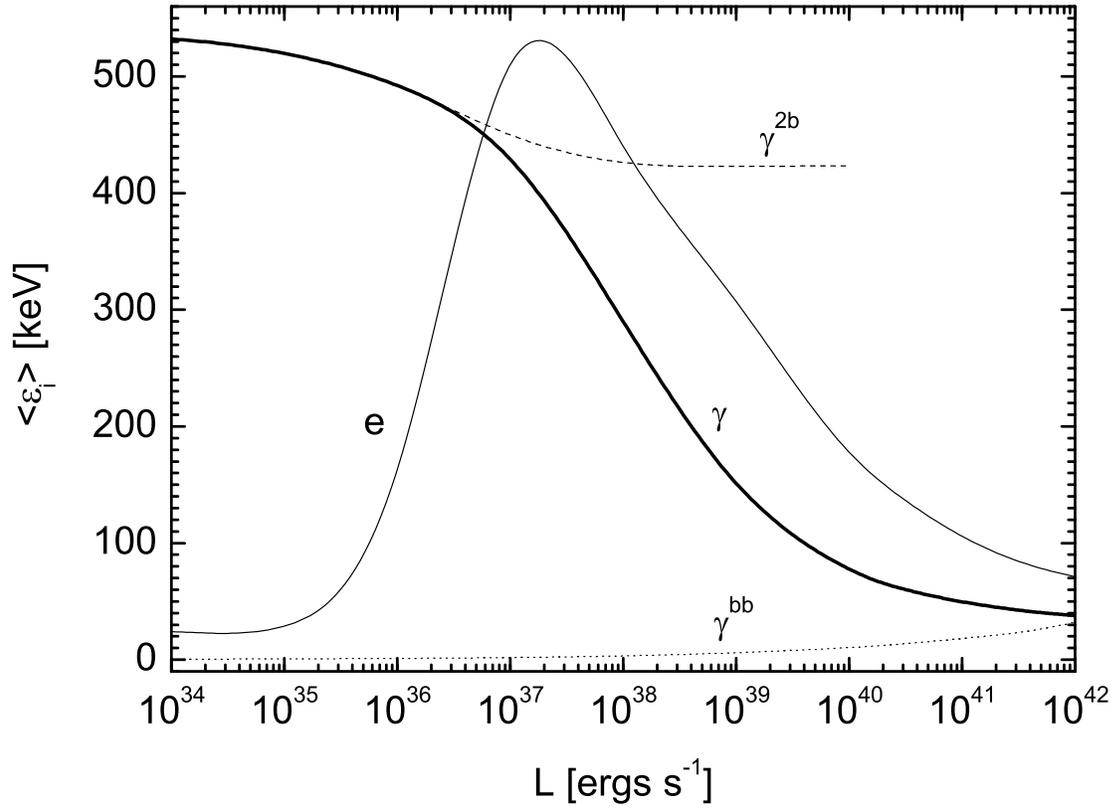} \caption{The mean energy of the emerging photons
(thick solid line) and electrons (thin solid line) as a function
of the total luminosity. For comparison, we show as the dotted
line the mean energy of blackbody photons for the same energy
density as that of the photons at the photosphere. Also shown as
the dashed line is the mean energies of the emerging photons in
the case when only two particle processes are taken into account.}
\label{epsilon_L.Fig}
\end{figure}

\begin{figure}
\plotone{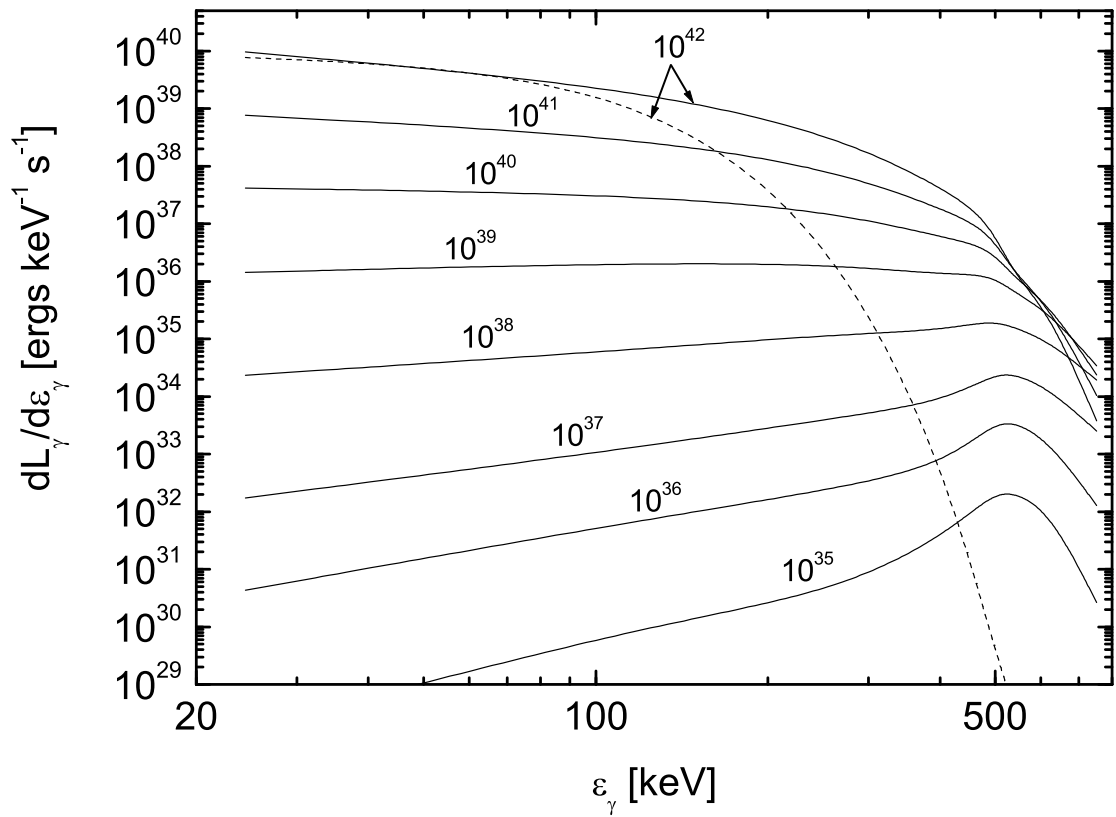} \caption{The energy spectrum of emerging photons
for different values of $\dot{\rm E}$, as marked on the curves.
The dashed line is the spectrum of blackbody emission.}
\label{spectra_gamma.Fig}
\end{figure}
\begin{figure}
\plotone{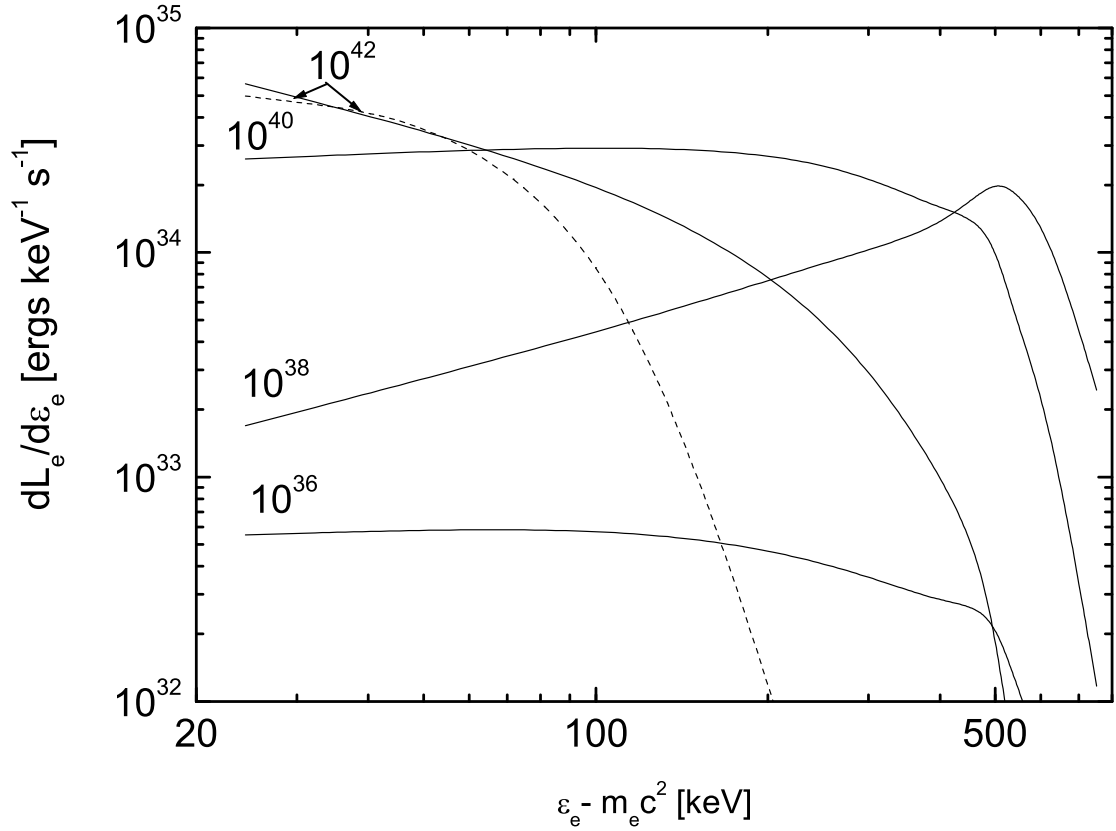} \caption{The energy spectrum of emerging pairs
for different values of $\dot{\rm E}$, as marked on the curves.
For comparison, the dashed line is the energy spectrum of pairs
that move with the velocity $v_e^{\rm out}\simeq 4.6\times 10^9$
cm~s$^{-1}$ and have a Maxwellian spectrum with a temperature of
$1.7\times 10^8$ K in the pair plasma frame.}
\label{spectra_e.Fig}
\end{figure}

\end{document}